\documentclass[aps,prd,reprint,amsmath,amssymb,superscriptaddress]{revtex4-1}
\usepackage{bbm}
\usepackage{bm}
\usepackage{graphicx}
\usepackage{mathrsfs}
\usepackage[pdftex,colorlinks,hyperindex,plainpages=false,bookmarksopen,bookmarksnumbered,pdfusetitle]{hyperref}
\usepackage{slashed}

\newcommand\muB{\mu_\text{B}}
\newcommand\dd{\mathrm{d}}
\newcommand\imag{\mathrm{i}}
\newcommand\gr[1]{\mathrm{#1}}
\newcommand\vek[1]{\bm{#1}}
\newcommand\threeint[1]{\int\frac{\dd^3\vek{#1}}{(2\pi)^3}}

\begin{document}

\title{Confronting effective models for deconfinement\\ in dense quark matter with lattice data}

\author{Jens O.~Andersen}
\email{jens.andersen@ntnu.no}
\affiliation{Department of Physics, Norwegian University of Science and Technology, H\o gskoleringen 5, 7491 Trondheim, Norway}

\author{Tom\'{a}\v{s} Brauner}
\email{tomas.brauner@uis.no}
\affiliation{Faculty of Science and Technology, University of Stavanger, 4036 Stavanger, Norway}

\author{William Naylor}
\email{william.naylor@ntnu.no}
\affiliation{Department of Physics, Norwegian University of Science and Technology, H\o gskoleringen 5, 7491 Trondheim, Norway}


\begin{abstract}
Ab initio numerical simulations of the thermodynamics of dense quark matter remain a challenge. Apart from the infamous sign problem, lattice methods have to deal with finite volume and discretization effects as well as with the necessity to introduce sources for symmetry-breaking order parameters. We study these artifacts in the Polyakov-loop-extended Nambu-Jona-Lasinio model, and compare its predictions to existing lattice data for cold and dense two-color matter with two flavors of Wilson quarks. To achieve even qualitative agreement with lattice data \emph{requires} the introduction of two novel elements in the model: (i) explicit chiral symmetry breaking in the effective contact four-fermion interaction, referred to as the chiral twist, and  (ii) renormalization of the Polyakov loop. The feedback of the dense medium to the gauge sector is modeled by a chemical-potential-dependent scale in the Polyakov-loop potential. In contrast to previously used analytical ans\"atze, we determine its dependence on the chemical potential from lattice data for the expectation value of the Polyakov loop. Finally, we propose to add a two-derivative operator to our effective model. This term acts as an additional source of explicit chiral symmetry breaking, mimicking an analogous term in the lattice Wilson action.
\end{abstract}


\maketitle


\section{Introduction}
\label{sec:intro}

Hadronic matter in extreme conditions such as high temperature or high density has received considerable attention over the past decades. However, direct numerical simulations of the theory of strong interactions --- quantum chromodynamics (QCD) --- at nonzero baryon density are a formidable challenge due to the infamous sign problem. Large efforts have been made to overcome this problem (see for instance Ref.~\cite{Cristoforetti:2012su,*Aarts:2013bla,*Sexty:2014zya} for some recent proposals), yet with limited success so far. At present, the only tools for quantitative analysis of dense nuclear matter are phenomenological effective models and, to some extent, continuum functional methods~\cite{Braun:2009gm,*Fischer:2011mz}.

Lattice simulations and phenomenological models can be of mutual benefit: while numerical simulations can provide a firm model-independent basis for effective continuum approaches, results obtained by the latter can be easily extrapolated to conditions where lattice techniques are difficult to apply. However, there is a gap that needs to be bridged to make this interaction possible. Lattice simulations have to deal with several artifacts, namely the effects of finite volume and spacetime discretization, as well as the need for external sources to pick a unique ground state whenever continuous symmetries are expected to be spontaneously broken. Usually, it is much easier to introduce these effects in models, rather than eliminate them from lattice simulations.

The main goal of the present paper is to do exactly that. We effectively re-introduce the effects of external sources and spacetime discretization using a phenomenological continuum model and focus on discriminating between physics and lattice artifacts. Our work is based upon results of recent simulations of {two-color} QCD (2cQCD) at high baryon density, using two flavors of Wilson-type quarks~\cite{Cotter:2012mb,Boz:2013rca}. These are compared to an effective model of the Nambu-Jona-Lasinio (NJL) type~\cite{Vogl:1991qt,Klevansky:1992qe,Buballa:2003qv}, augmented with the Polyakov loop, which is an (approximate) order parameter for deconfinement~\cite{Fukushima:2003fw,Ratti:2005jh,Megias:2004hj}. We discuss in detail the model under the constraints of spacetime and internal symmetries. Starting from a classification of all operators allowed in the Lagrangian by symmetries, we propose two modifications:
\begin{itemize}
\item Incorporation of explicit chiral symmetry breaking in the four-fermion interaction.
\item Addition of a two-derivative kinetic term, mimicking the lattice Wilson term.
\end{itemize}
We then show that the modification of the four-fermion interaction is \emph{required} for the model predictions to be consistent with lattice results, at least
away from the continuum limit (at fixed lattice spacing).

We would like to stress that it is not our purpose here to carry out a precise numerical fit of the effective model to available lattice data. We rather wish to gain qualitative insight with a reasonable number of free parameters, and thus to prepare the ground for a future more detailed quantitative study. The simple setting used here allows us to study separately, and make robust conclusions about, the various physical ingredients entering the problem, namely the chiral restoration at high density and low temperature as well as deconfinement at high temperature and moderate-to-high density.

The outline of the paper is as follows. In section~\ref{sec:2color} we review the most important properties of 2cQCD (see Ref.~\cite{vonSmekal:2012vx} for a recent review). Section~\ref{sec:model} is devoted to a detailed discussion of the model. We classify all operators up to dimension six, allowed by the symmetries, and give a brief overview of the qualitative changes brought 
about by the non-standard operators. Since we modify one of the interaction terms, it is mandatory to first analyze how it affects the physics in the vacuum. This is done in section~\ref{sec:vacuum}, where we also fix the parameters in the quark (NJL) sector of the model. The next two sections constitute the core of the paper, where we investigate various lattice artifacts in accordance with our main program. In section~\ref{sec:zeroT}, we work at zero temperature, allowing us to separate the physics of flavor symmetry from deconfinement issues and the Polyakov loop. We analyze in turn the effects of a diquark source, the dependence on the quark mass, and the modified interaction term. In section~\ref{sec:deconfinement}, we add the gauge sector to the model and study the thermodynamics at nonzero temperature; the renormalization of the Polyakov loop is a new element here. Subsequently, we extract the chemical potential dependence of the temperature scale in the Polyakov-loop potential, which essentially determines the position of the deconfinement crossover. In section~\ref{sec:conclusions} we summarize and conclude. Some calculation details can be found in appendix~\ref{app:wilson}, where we analyze in detail the consequences of the two-derivative (Wilson) kinetic term for quarks.


\section{Two-color QCD}
\label{sec:2color}

In this paper, 2cQCD means a non-Abelian gauge theory with the $\gr{SU(2)}$ gauge group and $N_f$ degenerate flavors of fundamental quarks. Two-color QCD differs in many respects from real-world, three-color QCD, most of them stemming from the fact that the fundamental representation of $\gr{SU(2)}$ is pseudoreal. For instance, a color singlet can only be made out of an even number of quarks, hence baryons in 2cQCD are bosons. Consequently, \emph{dilute} nuclear matter is expected to be formed by a Bose-Einstein condensate (BEC) of bosonic baryons rather than by a Fermi sea of nucleons.

Next comes the question of the low-energy hadronic spectrum, which is of importance for the thermodynamics at low temperatures. The lowest-lying states in the spectrum are determined by the spontaneously broken flavor symmetry in the (2c)QCD vacuum. Here the pseudoreality of quarks implies that the usual $\gr{SU}(N_f)_\gr{L}\times\gr{SU}(N_f)_\gr{R}\times\gr{U(1)_B}$ chiral symmetry of QCD in the limit of vanishing quark masses is embedded in an extended $\gr{SU}(2N_f)$ flavor symmetry group~\cite{Kogut:1999iv,*Kogut:2000ek}. The chiral condensate in the vacuum breaks this to $\gr{Sp}(2N_f)$, leading to $2N_f^2-N_f-1$ Goldstone bosons. These include $N_f^2-1$ pseudoscalar mesons, and $N_f(N_f-1)/2$ diquark-antidiquark pairs. Nonzero (degenerate) quark masses make these modes massive. Forming an irreducible multiplet of the $\gr{Sp}(2N_f)$ symmetry, these states are all degenerate with a common mass that we denote by $m_\pi$.

The determinant of the Dirac operator of 2cQCD is necessarily real and for an even number of degenerate quark flavors it is also positive~\cite{Hands:2000ei}. Consequently, the theory does not suffer from the sign problem and Monte Carlo simulations of dense matter are possible. In the following, we will focus exclusively on the simplest case of two quark flavors. In this case, the usual pion triplet is augmented by a single diquark-antidiquark pair carrying baryon number but no isospin. This determines the basic topology of the phase diagram of 2cQCD. Nonzero baryon chemical potential $\muB$ breaks the flavor symmetry down to the usual $\gr{SU}(2)_\gr{L}\times\gr{SU}(2)_\gr{R}\times\gr{U(1)_B}$ chiral group; this is natural as the additional symmetry generators following from the pseudoreality of the quark representation do not commute with the baryon number operator. For nonzero (degenerate) quark masses, only the $\gr{SU}(2)_\gr{V}\times\gr{U(1)_B}$ subgroup is exact.

When $\muB\geq m_\pi$, the diquarks are expected to undergo BEC, which breaks $\gr{U(1)_B}$ spontaneously. Since $\gr{U(1)_B}$ is an exact symmetry, the baryon superfluid phase is necessarily separated from the vacuum by a phase transition. As the chemical potential is further increased, one eventually enters a Bardeen-Cooper-Schrieffer-like (BCS-like) regime where the thermodynamics is dominated by a Fermi sea of quarks that form weakly bound Cooper pairs. The order parameter for $\gr{U(1)_B}$ symmetry breaking now is a composite diquark operator, which has the same quantum numbers as the order parameter in the BEC phase. The two regimes should therefore be smoothly connected~\cite{Son:2000xc}; in this context one speaks of a BEC-BCS crossover~\footnote{Here we identify the position of the BEC-BCS crossover with the region where the quark chemical potential measured with respect to its constituent mass changes sign. Another useful measure of the crossover is provided by the diquark wave function, recently studied in 2cQCD in Ref.~\cite{Amato:2015gea}.}.

If we instead crank up the temperature at zero baryon chemical potential, we expect the physics to be more-or-less similar to that of three-color QCD. Around some pseudocritical temperature $T_c$, we expect a rapid crossover from hadronic to quark degrees of freedom, accompanied by a rise in the expectation value of the Polyakov loop. This is loosely referred to as deconfinement. In the same range of temperatures, the chiral condensate melts, and we 
enter the quark-gluon plasma phase.

In lattice simulations of 2cQCD with Wilson fermions, the symmetry of the theory is affected in a twofold manner. Firstly, Lorentz invariance is broken to a discrete symmetry group of the spacetime lattice, which may result in a certain degree of anisotropy. At the same time, the Wilson term in the action acts as an additional source of explicit breaking of flavor symmetry which only disappears in the continuum limit. Since it breaks the flavor symmetry in exactly the same way as the quark masses do \emph{it may not be possible to distinguish the two sources of symmetry breaking in low-energy observables}.

In our model calculations, we will focus on two classes of observables: symmetry-breaking order parameters (condensates) and thermodynamic quantities such as pressure or baryon density. We now summarize the expectations for these observables and then describe the relevant lattice results in order to set the stage for our analysis.


\subsection{Chiral perturbation theory}
\label{subsec:CHPT}

The flavor symmetry of 2cQCD and its spontaneous breaking are most conveniently encoded in the low-energy effective theory for the pseudo-Goldstone modes, namely chiral perturbation theory ($\chi$PT)~\cite{Gasser:1983yg,*Gasser:1984gg}. Following the notation of Refs.~\cite{Kogut:2000ek,Brauner:2006dv}, the pions and diquarks are expressed in terms of a $2N_f\times2N_f$ antisymmetric unimodular unitary matrix $\Sigma$, in terms of which the leading-order $\chi$PT Lagrangian in Minkowski space reads
\begin{equation}
\mathscr L_{\chi\text{PT}}=\frac12f_{\pi}^2\,\text{tr}(D_\mu\Sigma D^\mu\Sigma^\dagger)+H\,\text{Re}\,\text{tr}(J\Sigma).
\label{LagrangianCHPT}
\end{equation}
Here $f_\pi$ is the pion decay constant and the covariant derivative $D_\mu$ includes the baryon number chemical potential $\muB$, $D_\mu\Sigma\equiv\partial_\mu\Sigma-\imag\delta_{\mu0}\muB(B\Sigma+\Sigma B^T)$, where $B$ is the baryon number operator. The matrix $J$ in general contains sources that couple to scalar or pseudoscalar quark bilinears. Here we will need the scalar source $m_0$ (quark mass), and the diquark source $j$, in terms of which we have $J=m_0\Sigma_1^\dagger+j\Sigma_2^\dagger$. In the case of two flavors, the matrix basis can be chosen such that
\begin{equation}
B=\frac12
\begin{pmatrix}
\mathbbm1 & 0\\
0 & -\mathbbm1
\end{pmatrix},
\hspace{0.1cm}
\Sigma_1=
\begin{pmatrix}
0 & -\mathbbm1\\
\mathbbm1 & 0
\end{pmatrix},
\hspace{0.1cm}
\Sigma_2=
\begin{pmatrix}
\tau_2 & 0\\
0 & \tau_2
\end{pmatrix}
\end{equation}
in the $2\times2$ block form, where $\tau_2$ is a Pauli matrix. Finally, the coupling $H$ can be fixed by expanding the Lagrangian~\eqref{LagrangianCHPT} to second order in the fluctuations about the vacuum, or by using the Gell-Mann-Oakes-Renner relation, leading to $H=f_\pi^2m_\pi^2/m_0$.

As the low-energy spectrum contains excitations carrying baryon number, the leading-order chiral Lagrangian~\eqref{LagrangianCHPT} can be used to study the phase diagram of 2cQCD at zero temperature and nonzero baryon density~\footnote{Including the effects of nonzero temperature requires calculating loops, and hence going to the next-to-leading order in the derivative expansion~\cite{Splittorff:2001fy,*Splittorff:2002xn}.}. The ground state at finite density is most easily visualized by using a Lie algebra isomorphism to cast the coset space $\gr{SU(4)/Sp(4)}$ equivalently as $\gr{SO(6)/SO(5)}$~\cite{Brauner:2006dv}. The unitary matrix $\Sigma$ can thus be mapped onto a unit 6-vector $\vec n$ via the relation $\Sigma=\vec n\cdot\vec\Sigma$, where $\Sigma_i$ is a set of suitably chosen basis matrices. Depending on the values of the chemical potential(s) and the external sources, the ground state therefore moves on a unit sphere. In the absence of an isospin chemical potential $\mu_\text{I}$ and other sources except $m_0,j$, the symmetry of the problem can be exploited to rotate the ground state into the $(n_1,n_2)$ plane, corresponding to a chiral condensate and a diquark condensate with a fixed phase. The ground state can thus be parametrized by a single angle $\theta$ such that $n_1=\cos\theta$ and $n_2=\sin\theta$.

For the time being, we will assume that $j=0$. The effects of the diquark source will be discussed in detail in section~\ref{subsec:source}. The static part of the Lagrangian~\eqref{LagrangianCHPT}, whose \emph{maximum} is to be found, then becomes
\begin{equation}
\mathscr L_{\chi\text{PT}}^\text{stat}=2f_\pi^2\muB^2\sin^2\theta+4f_\pi^2m_\pi^2
\cos\theta.
\label{LCHPTstat}
\end{equation}
As expected, the chiral-symmetry-breaking state $\theta=0$ is stable for $\muB<m_\pi$. As $\muB$ further increases, the equilibrium starts rotating into the diquark direction, and the angle of rotation $\theta$ is given by~\cite{Kogut:2000ek}
\begin{equation}
\cos\theta=\frac{m_\pi^2}{\muB^2}.
\label{vacuumangle}
\end{equation}
This result is a priori expected to hold only within the range of validity of $\chi$PT, in particular only in the BEC regime where bosonic degrees of freedom dominate the physics of dense two-color matter. It should therefore be emphasized that, in fact, it remains at least qualitatively accurate even for much higher values of $\muB$. A numerical solution of the NJL model shows that Eq.~\eqref{vacuumangle} holds also for chemical potentials rather deep in the BCS phase where a Fermi sea of quarks has been formed. The agreement between the two approaches, based on completely different degrees of freedom, persists at zero temperature up to $\muB\approx3m_\pi$~\cite{Andersen:2010vu}. This can be attributed to the fact that in this range of $\muB$, the physics at zero temperature is almost entirely driven by the condensates; the contribution of quark quasiparticles is negligible.


\subsection{The puzzle and its resolution}
\label{subsec:puzzle}

The above discussion hints that the quantitative aspects of the phase diagram of 2cQCD are determined by its symmetries even far beyond the region where one would expect it. The results of lattice simulations are in a stark contrast to this naive expectation. Most importantly, the lattice data indicate a fast transition to a BCS regime just above $\muB=m_\pi$; the expected bosonic BEC phase, if present at all, is not resolved~\footnote{Note that this strong statement only applies to the recent simulations with Wilson-type quarks. The results of previous simulations using \emph{staggered} quarks~\cite{Hands:2000ei,Hands:2001ee}, concerning the values of the chiral and the diquark condensate as a function of baryon number chemical potential, were in agreement with $\chi$PT. These data alone, however, do not allow one to distinguish between the BEC and BCS regimes~\cite{Andersen:2010vu}.}. Moreover, in the BCS regime, 2cQCD behaves as a system of weakly interacting, almost massless quarks. This conclusion is supported by two independent pieces of evidence~\cite{Cotter:2012mb}:
\begin{itemize}
\item The values of pressure and baryon number density, when normalized to their values for an ideal gas of massless quarks --- the Stefan-Boltzmann (SB) limit --- exhibit a plateau at $\muB\gtrsim m_\pi$ close to one. The precise height of this plateau is currently hard to determine, but its existence seems to be confirmed.
\item The expectation value of the color-singlet diquark operator scales with $\muB^2$ in the same range of chemical potentials, reminiscent of the density of states at a sharp Fermi surface of massless relativistic fermions.
\end{itemize}
An additional, related piece of evidence is provided by the fact that the critical temperature $T_d$ for diquark condensation saturates at high $\muB$ at a value roughly given by $T_d\approx T_c/2$~\cite{Boz:2013rca}, whereas the picture of the order parameter rotating on a unit sphere, sketched above, would naively suggest a vastly different value, $T_d\approx T_c$.

How can we reconcile these observations with the universal model-independent predictions of $\chi$PT? The simple and short answer is: we \emph{cannot}. The SB scaling of thermodynamic observables requires that the quarks are almost massless and their quasiparticle gap, proportional to the diquark condensate, is very small as well. It is obvious that this cannot be true simultaneously in $\chi$PT, according to which the sum of squares of the chiral and diquark condensates is constant. 

It is apparent that the key physical ingredient required is the rapid transition to a BCS-like gas of almost massless and gapless quarks at $\muB\gtrsim m_\pi$. Once this is achieved, the other features --- $\muB^2$-scaling of the expectation value of the diquark operator and the suppression of $T_d$ --- should follow naturally. $\chi$PT as well as effective models based on the same symmetries lead to predictions that are at odds with this requirement. We therefore expect explicit chiral symmetry breaking to play a crucial role. However, as we demonstrate in section~\ref{subsec:qmass}, tuning the current quark mass even to unreasonably high values is not sufficient. In the remainder of this paper, we therefore take a rather radical approach to the problem 
in which we abandon the chiral symmetry altogether. We introduce explicit symmetry breaking into the effective four-quark interaction and show that this leads to the desired effect. The possible origin of this symmetry breaking is discussed in section~\ref{sec:conclusions}.


\section{The model}
\label{sec:model}

Our model is of PNJL type, which is based on quark degrees of freedom. Thus the only field variable is the quark spinor $\psi$. For the time being, we focus on the quark sector. The gauge sector will be discussed in detail later, when the effects of nonzero temperature are introduced. The form of the Lagrangian is constrained by spacetime and internal symmetries, which we assume to be as follows: Poincar\'e invariance plus the discrete symmetries of charge conjugation, parity, and time reversal, \emph{global} $\gr{SU(2)}$ color symmetry, and $\gr{Sp}(4)$ flavor symmetry. We therefore abandon the full  $\gr{SU}(4)$ flavor group. This group is explicitly broken down to the $\gr{Sp}(4)$ subgroup by (degenerate) quark masses and the Wilson term in the lattice action. The $\gr{Sp}(4)$ subgroup therefore constitutes the true flavor symmetry of both lattice and continuum 2cQCD with massive quarks.

Finally, we make one more step which goes beyond the usual PNJL model building. We do not restrict ourselves to the simplest possible Lagrangian consisting of a quark kinetic term and a four-quark interaction. Instead, we classify all terms in the Lagrangian consistent with the above symmetries up to dimension six~\footnote{Dimension-six operators are needed to describe quark interactions.}. This is in agreement with the effective field theory philosophy where all operators up to a given dimension allowed by symmetry should be included~\cite{Weinberg:1995v1}.


\subsection{Classification of operators}
\label{subsec:classification}

We start by classifying the operators according to their canonical dimension. The operators are written schematically with their indices suppressed.
\begin{itemize}
\item\textbf{Order 3.} The only parity-even Lorentz scalar that respects baryon number, $\gr{SU(2)}$ color and isospin invariance is the quark mass term $\bar\psi\psi$. The $\gr{Sp}(4)$ symmetry is automatically implied, but the full $\gr{SU}(4)$ group is explicitly broken. 
\item\textbf{Order 4.} Schematically, the operator must be of the $\bar\psi D\psi$ type to respect baryon number. Lorentz invariance and parity together with the $\gr{SU(2)}$ color and isospin invariance single out the usual quark kinetic term $\bar\psi\slashed D\psi$ . The full $\gr{SU}(4)$ symmetry is automatically implied. 
\item\textbf{Order 5.} Here we have two possibilities respecting baryon number conservation and Lorentz invariance, namely $D_\mu\bar\psi D^\mu\psi$ and $D_\mu\bar\psi[\gamma^\mu,\gamma^\nu]D_\nu\psi$. The latter is, however, irrelevant for our purposes since we only consider a purely temporal background gauge field, representing the baryon chemical potential and the Polyakov loop. Parity, $\gr{SU(2)}$ color and isospin invariance then single out the operator $D_\mu\bar\psi D^\mu\psi$. The $\gr{Sp}(4)$ symmetry is automatically implied, but the full $\gr{SU}(4)$ group is explicitly broken. We will refer to this operator as the \emph{Wilson term} since it closely resembles the corresponding operator in the lattice Wilson action.
\item\textbf{Order 6.} Here we have two schematic possibilities respecting baryon number, namely $\bar\psi DDD\psi$ and $(\bar\psi\psi)^2$. Lorentz invariance requires the former to contain an odd number of Dirac matrices, hence it represents a higher-order correction to the kinetic term which automatically preserves chiral symmetry. We will therefore drop it from further consideration since we do not expect that it leads to any qualitatively new effects. The non-derivative operator, on the other hand, is a standard NJL-type four-quark interaction. There are many operators of this type  that respect a given symmetry, as can be seen by performing a Fierz transformation of the basic one-gluon-exchange type of operator~\cite{Buballa:2003qv}. In the mean-field approximation, we want to build an invariant interaction out of fermion bilinears that carry the quantum numbers of the low-energy degrees of freedom. The degrees of freedom that must be present in the model are the pseudo-Goldstone bosons: the pion triplet (represented by $\bar\psi\imag\gamma_5\vec\tau\psi$) and the isospin-singlet diquark, which is represented by $\bar\psi^{\mathcal C}\gamma_5\sigma_2\tau_2\psi$ (where $\mathcal C$ stands for charge conjugation and $\sigma_2$ is the Pauli matrix in color space). Furthermore, the operator with the quantum numbers of a true scalar, $\bar\psi\psi$, must be added in order to account for the chiral condensate. The two $\gr{Sp}(4)$-invariant interactions that can be built out of the squares of these operators are
\end{itemize}
\begin{equation}
(\bar\psi\psi)^2\quad\text{and}\quad(\bar\psi\imag\gamma_5\vec\tau\psi)^2+|\bar\psi^{\mathcal C}\gamma_5\sigma_2\tau_2\psi|^2.
\end{equation}
The above considerations suggest the following generic NJL-type Lagrangian,
\begin{equation}
\begin{split}
\mathscr L_\text{NJL}={}&\bar\psi(\imag\slashed D-m_0)\psi+\kappa D_\mu\bar\psi D^\mu\psi+G(\bar\psi\psi)^2\\
&+\lambda G\left[(\bar\psi\imag\gamma_5\vec\tau\psi)^2+|\bar\psi^{\mathcal C}\gamma_5\sigma_2\tau_2\psi|^2\right],
\end{split}
\label{LagrangianNJL}
\end{equation}
where the covariant derivative $D_\mu\psi$ includes the baryon chemical potential, and later in section~\ref{sec:deconfinement} also the constant background gauge field representing the Polyakov loop. The dimensionless parameter $\lambda$ is in the remainder of this paper referred to as the \emph{chiral twist}. The minimal NJL model for three-color QCD amounts to skipping the last operator inside the brackets, and setting $\kappa=0$ and $\lambda=1$~\cite{Klevansky:1992qe}. The two-color version including the diquark channel, with $\kappa=0$ and $\lambda=1$, was first introduced in Ref.~\cite{Ratti:2004ra} and subsequently used by several other groups~\cite{Brauner:2009gu,Sun:2007fc,*Imai:2012hr,*Makiyama:2015uwa}. In the following subsection, we will discuss the consequences of the modifications introduced here.


\subsection{Effects of chiral symmetry breaking}
\label{subsec:modelsetup}

The classification of invariant operators has automatically guided us to introduce two new couplings in the Lagrangian. Let us start with the Wilson term proportional to $\kappa$. As already mentioned, it mimics the discretization artifacts introduced by the lattice Wilson action. The presence of a two-derivative bilinear operator in the Lagrangian leads to doubling of the fermion degrees of freedom. For small $\kappa$, we expect the new fermion species to be heavy with mass scaling as $1/\kappa$. Since the Wilson term breaks chiral symmetry explicitly, it will enhance the chiral condensate in the vacuum. Regarding thermodynamics, we expect the effects of the Wilson term to be most pronounced at high temperatures since the new fermion is heavy and thus difficult to excite thermally. In addition, the presence of a heavy fermion may lead to undesirable artifacts at high baryon density, namely the appearance of a second Fermi surface at high $\muB$ and low temperature. Some of the quantitative consequences of the Wilson term will be worked out in appendix~\ref{app:wilson}.

It is the chiral twist that will play a pivotal role in our analysis. As we abandon chiral symmetry here, the scalar and pseudoscalar channels are no longer forced to appear with the same strength. However, the diquark and pion operators do enter through a fixed combination, as required by the exact $\gr{Sp}(4)$ symmetry of two-flavor 2cQCD. This guarantees exact mass degeneracy of pions and diquarks, and therefore that the critical chemical potential for diquark condensation at zero temperature equals the \emph{pion} mass.

A detailed analysis of the consequences of our model~\eqref{LagrangianNJL} with the modified interaction constitutes the bulk of the remainder of the paper. Yet some qualitative observations can be made already now. It is clear that the existence of a bound state in the pion channel requires $\lambda>0$. Since the strength of attraction in the scalar channel remains fixed to $G$, the constituent quark mass in the vacuum will be unaffected by $\lambda$. As a consequence, the pion will become less strongly bound and its mass will increase with decreasing $\lambda$. The location of the BEC-BCS crossover at zero temperature will therefore shift towards lower values of $\muB/m_\pi$ and it is conceivable that the BEC region can be eliminated altogether. In the following, we will therefore consider only the range $0<\lambda\leq1$.

Since the coupling in the diquark channel also decreases with $\lambda$, we expect the diquark condensate to be suppressed. This will sharpen the quark Fermi surface and move the system closer to the SB limit. An additional effect which follows from our analysis below, is that the chiral condensate will also be suppressed as compared to the naive expectation based on Eq.~\eqref{vacuumangle}. Suppressing both the diquark and the chiral condensate, the chiral twist is therefore exactly what we need in order to explain the rapid crossover to the gas of weakly coupled almost massless quarks.

From the strict symmetry point of view, the Wilson term and the chiral twist should be treated on the same footing as they break symmetry in the same way. However, a detailed numerical investigation reveals that the Wilson term has a rather marginal effect on the observables considered in this paper, and certainly cannot resolve the puzzle presented in section~\ref{subsec:puzzle}. We therefore take the liberty and discard the Wilson term altogether from most of our analysis. This step is driven by simplicity: as explained in detail in appendix~\ref{app:wilson}, including the Wilson term requires a substantial modification of the framework including a change of regularization scheme. At the level of accuracy, considered in this paper, the chiral twist turns out to be the only necessary and sufficient new ingredient in the model.


\section{Vacuum physics}
\label{sec:vacuum}

\begin{figure*}
\includegraphics[height=0.37\textwidth]{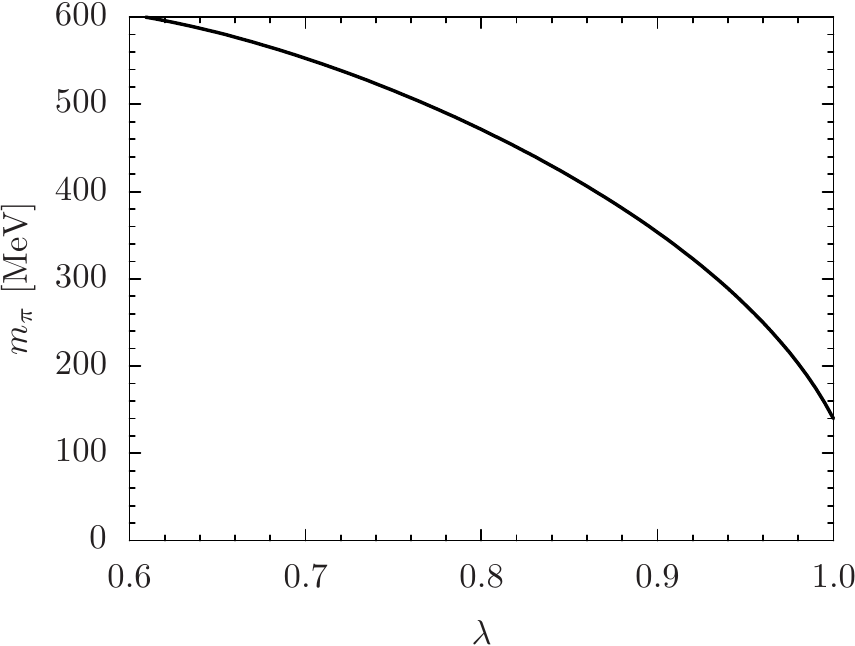}
\hfill
\includegraphics[height=0.37\textwidth]{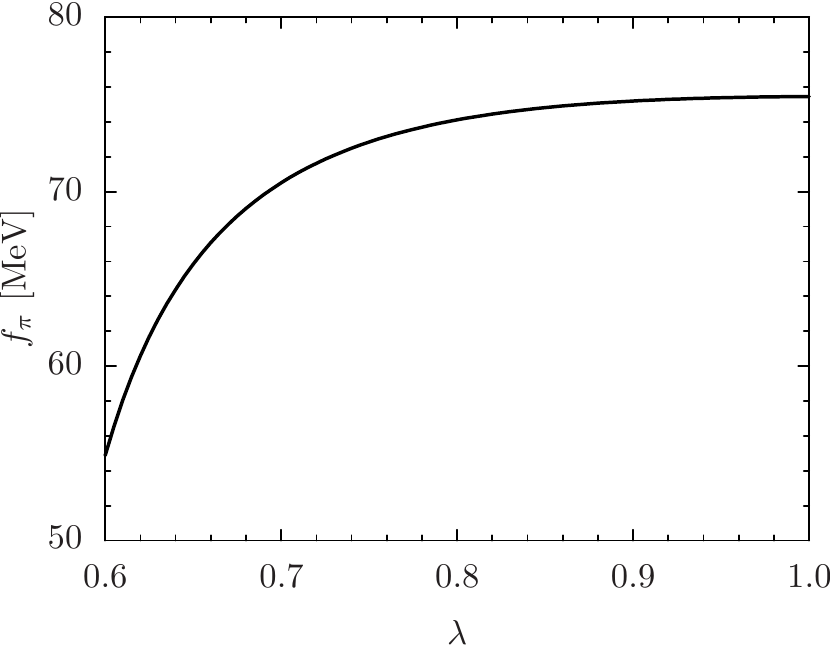}
\caption{Pion mass (left panel) and decay constant (right panel) as a function of $\lambda$. All other parameters are fixed according to Eq.~\eqref{parameters}.}
\label{fig:mpifpi}
\end{figure*}

In the remainder of the paper, except for appendix~\ref{app:wilson}, we set $\kappa=0$. The composite bosonic degrees of freedom $\sigma$, $\vec\pi$, $\Delta$, and $\Delta^*$ are introduced by the Hubbard-Stratonovich transformation, adding to the Lagrangian~\eqref{LagrangianNJL} the term
\begin{align}
\notag
\Delta\mathscr L_\text{NJL}={}&-\frac1{4G}(\sigma+2G\bar\psi\psi)^2-\frac1{4\lambda G}(\vec\pi+2\lambda G\bar\psi\imag\gamma_5\vec\tau\psi)^2\\
&-\frac1{4\lambda G}|\Delta^*-2\lambda G\bar\psi\imag\gamma_5\sigma_2\tau_2\psi^{\mathcal C}|^2.
\end{align}
The Lagrangian is now bilinear in the quark fields and we can integrate them out exactly. This leads to the effective bosonic action
\begin{equation}
S^\text{eff}_\text{NJL}=-\int\dd t\,\dd^3\vek x\left(\frac{\sigma^2}{4G}+\frac{\vec\pi^2+|\Delta|^2}{4\lambda G}\right)-\imag\,\text{Tr}\log\mathscr D,
\label{master_action}
\end{equation}
where
\begin{equation}
\mathscr D\equiv\begin{pmatrix}
\imag\slashed D-M-\imag\gamma_5\vec\pi\cdot\vec\tau & \Delta\gamma_5\\
-\Delta^*\gamma_5 & \imag\slashed D-M+\imag\gamma_5\vec\pi\cdot\vec\tau
\end{pmatrix}
\end{equation}
is the Dirac operator acting on the Nambu space spanned by red quarks and green antiquarks. We have introduced the shorthand notation $M\equiv m_0+\sigma$ for the constituent quark mass, and denoted by ``$\text{Tr}$'' the functional trace.

In the following, we work in the mean-field approximation. This means that after taking functional derivatives of the action~\eqref{master_action} as appropriate, all bosonic fields are set equal to a constant. The integrals that appear are ultraviolet divergent and we need to regulate them. With the exception of appendix~\ref{app:wilson}, we will use a simple sharp three-momentum cutoff $\Lambda$. Depending on the cutoff, writing the expressions in a manifestly Lorentz-covariant form may involve some manipulations that are not justified from a strictly mathematical point of view. This is, however, a well-known issue~\cite{Klevansky:1992qe}. 

The vacuum of 2cQCD is characterized by broken chiral symmetry. Differentiating Eq.~\eqref{master_action} with respect to $m_0$, we find that the chiral condensate of a single quark flavor is related to the $\sigma$ condensate by
\begin{equation}
\langle\bar uu\rangle=-\frac\sigma{4G}.
\label{condchi}
\end{equation}
The latter is found from the stationary point condition $\delta S^\text{eff}_\text{NJL}/\delta\sigma=0$. This yields~\footnote{We keep the dependence of the analytic formulas on the number of colors $N_c$ explicit in order to facilitate a comparison with the three-color case.}
\begin{equation}
\sigma=16\imag GN_cM\int\frac{\dd^4 k}{(2\pi)^4}\frac 1{k^2-M^2}=8GN_cM\threeint k\frac1{\epsilon_{\vek k}},
\label{condsigma}
\end{equation}
where the quark dispersion relation in the vacuum is
\begin{equation}
\epsilon_{\vek k}\equiv\sqrt{\vek k^2+M^2}.
\end{equation}
The spectrum of pseudo-Goldstone bosons can be determined from the polarization function (inverse propagator), which is obtained by taking a second functional derivative of the action. We already know that in the vacuum pions and diquarks are degenerate, so we just state the result for the pion polarization function, simplified by using Eq.~\eqref{condsigma},
\begin{equation}
\chi(p^2)=-\frac1{2G}\left(\frac1\lambda-\frac\sigma M\right)+4N_cp^2I(p^2),
\end{equation}
where
\begin{equation}
I(p^2)\equiv-\imag\int\frac{\dd^4 k}{(2\pi)^4}\frac1{[(k+p)^2-M^2](k^2-M^2)}.
\label{pionpolarization}
\end{equation}
In the chiral limit, $M=\sigma$ and we can immediately see the effect of explicit chiral symmetry breaking by the chiral twist: the pion is exactly massless only for $\lambda=1$. In general the pion mass squared is given by the zero of the inverse propagator, $\chi(m_\pi^2)=0$. Using the the gap equation~\eqref{condsigma} once more, we can rewrite this as
\begin{equation}
1=8\lambda GN_c\threeint k\left(\frac1{2\epsilon_{\vek k}+m_\pi}+\frac1{2\epsilon_{\vek k}-m_\pi}\right).
\label{pionpole}
\end{equation}
Using the Lehmann spectral representation of the pion propagator, we can determine the coupling $g_{\pi qq}$ of the one-pion state to the pseudoscalar quark bilinear $\bar\psi\imag\gamma_5\vec\tau\psi$,
\begin{equation}
\frac1{g_{\pi qq}^2}=\chi'(p^2)\Bigr|_{p^2=m_\pi^2}=16N_c\threeint k\frac{\epsilon_{\vek k}}{(4\epsilon_{\vek k}^2-m_\pi^2)^2}.
\label{gpiqq}
\end{equation}
From this result, one can obtain the coupling of the one-pion state to the axial vector current, that is, the pion decay constant $f_\pi$. The resulting expression is
\begin{equation}
f_\pi=\frac{g_{\pi qq}}{2Gm_\pi^2}\left(\frac M\lambda-\sigma\right).
\label{fpi}
\end{equation}


\subsection{Parameter fixing}

The NJL model~\eqref{LagrangianNJL} with $\kappa=0$ and $\lambda=1$ is defined by the three parameters $m_0$, $G$, and $\Lambda$, which should be determined by a fit to three independent observables. It is customary to use the chiral condensate, pion decay constant and pion mass for that purpose. We follow Ref.~\cite{Brauner:2009gu} and determine the values of these input quantities in 2cQCD from their physical, three-color counterparts using a naive scaling with the number of colors $N_c$,
\begin{equation}
\begin{split}
\langle\bar uu\rangle&=-(218\text{ MeV})^3,\\
f_\pi&=75.4\text{ MeV},\qquad\text{(physical input)}\\
m_\pi&=140\text{ MeV}.
\end{split}
\label{observables}
\end{equation}
Equations~\eqref{condchi}, \eqref{pionpole}, and \eqref{fpi} then give us the following values
\begin{equation}
\begin{split}
G&=7.23\text{ GeV}^{-2},\\
\Lambda&=657\text{ MeV},\qquad\text{(fitted parameters)}\\
m_0&=5.4\text{ MeV},
\end{split}
\label{parameters}
\end{equation}
which we will use throughout the rest of the paper unless explicitly stated otherwise. The coupling $\lambda$ will be treated as a tunable parameter.


\subsection{Role of the chiral twist}

As explained in section~\ref{subsec:modelsetup}, we expect that the pion mass increases with decreasing $\lambda$. It is clear from the left panel of figure~\ref{fig:mpifpi} that the effect is actually rather large. At $\lambda\approx0.95$, the pion mass is increased by a factor of two, and at around $\lambda\approx0.6$, it reaches the (unphysical) threshold for decay into a quark-antiquark pair. (The $\sigma$ condensate in the vacuum is to great precision equal to $300\text{ MeV}$ for our choice of parameters.) At the same time, the pion decay constant starts to drop rapidly (right panel of figure~\ref{fig:mpifpi}), which indicates that the pion ceases to behave as a Goldstone boson. The interval $[0.6,1]$ therefore defines the range of reasonable values for $\lambda$ to which we will restrict ourselves in the following.

To gain a more analytic insight into the $\lambda$-dependence of physical observables, we take the derivative of Eq.~\eqref{pionpole} with respect to $\lambda$. Using Eq.~\eqref{gpiqq} allows us to rewrite the result in the form of an exact differential equation,
\begin{equation}
\frac{\dd m_\pi^2}{\dd\lambda}=-\frac{g_{\pi qq}^2}{2\lambda^2 G}.
\end{equation}
In the chiral limit, the pion mass will scale asymptotically as $m_\pi\propto\sqrt{1-\lambda}$ for $\lambda\to1$. Using in addition Eq.~\eqref{fpi}, we readily obtain another differential equation, this time for the pion decay constant,
\begin{equation}
f_\pi^2=\frac1{2G}(M-\lambda\sigma)^2\frac\dd{\dd\lambda}\frac1{m_\pi^2}.
\end{equation}
This expression together with the asymptotic scaling $m_\pi\propto\sqrt{1-\lambda}$ explains the very weak $\lambda$-dependence of $f_\pi$ for $\lambda$ close to one, see the right panel of figure~\ref{fig:mpifpi}. The fact that $f_\pi$ is almost constant in the range $\lambda\in[0.8,1]$ also justifies a posteriori our treatment of $\lambda$ as a tunable parameter: since the quark mass, and hence the pion mass, is a free parameter on the lattice anyway, we have some freedom in tuning both $m_0$ and $\lambda$ without affecting the physical observables $\langle\bar uu\rangle$ and $f_\pi$, and do not have to refit our parameters anew for each value of $\lambda$.


\section{Zero temperature: chiral restoration}
\label{sec:zeroT}

In this section, we will investigate the effects of tuning various parameters on different physical quantities at zero temperature. We focus on the chiral and diquark condensates, the pressure, and the baryon number density. We can therefore drop the pion field $\vec{\pi}$ in Eq.~\eqref{master_action}. Evaluating the action for constant $\sigma$ and $\Delta$ and going to Euclidean space gives the thermodynamic potential
\begin{equation}
\Omega^{T=0}_{\text{NJL}}=\frac{\sigma^2}{4G}+\frac{|\Delta|^2}{4\lambda G}-2N_c\sum_\pm\threeint k\,E^\pm_{\vek k},
\label{OmegaT0}
\end{equation}
where the quark quasiparticle dispersion relations are defined by~\footnote{For the quark degrees of freedom, we choose to work with the \emph{quark} number chemical potential $\mu$, given by $\mu=\muB/2$.}
\begin{equation}
E^\pm_{\vek k}\equiv\sqrt{(\xi^\pm_{\vek k})^2+|\Delta|^2},\qquad
\xi^\pm_{\vek k}\equiv\epsilon_{\vek k}\pm\mu.
\end{equation}
The values of the condensates for a given $\mu$ are found by direct numerical minimization of the thermodynamic potential. The pressure is equal to $-\Omega^{T=0}_{\text{NJL}}$ conventionally shifted by a constant so that the pressure is zero in the vacuum. The baryon number density is obtained by taking the derivative of the pressure with respect to $\muB$, giving
\begin{equation}
n_\text{B}^{T=0}=N_c\threeint k\left(\frac{\xi^+_{\vek k}}{E^+_{\vek k}}-\frac{\xi^-_{\vek k}}{E^-_{\vek k}}\right),
\label{Bdensity}
\end{equation}
where we have separated the particle and antiparticle contributions.


\subsection{Role of the diquark source}
\label{subsec:source}

\begin{figure}
\includegraphics[width=\columnwidth]{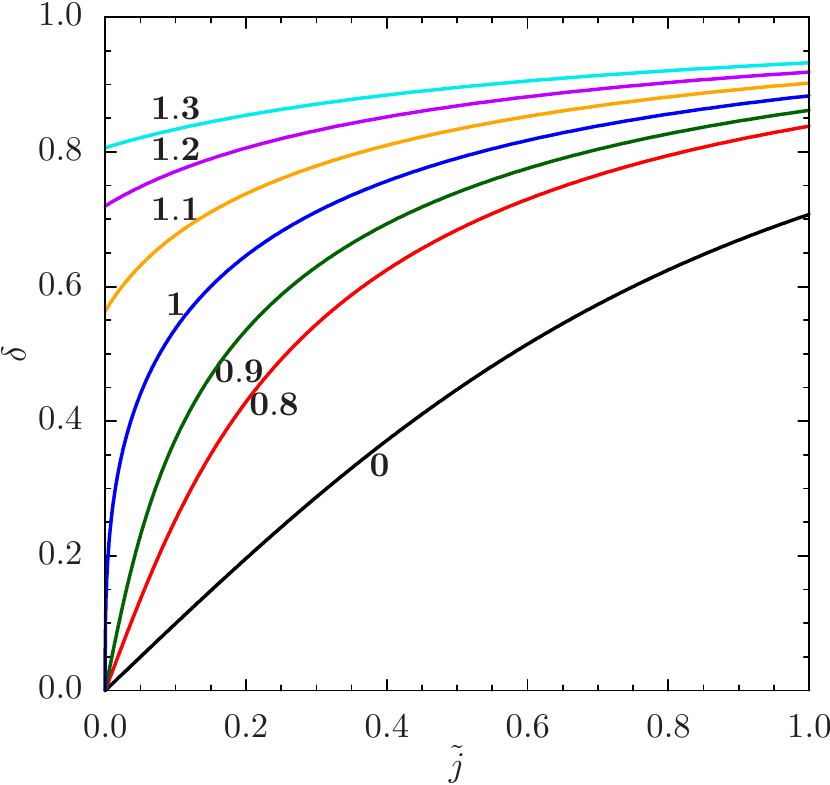}
\caption{Dependence of the rescaled diquark condensate $\delta$ on the dimensionless source $\tilde j$ for several different values of the parameter $x=\muB/m_\pi$ (given in bold).}
\label{fig:CHPTsource}
\end{figure}

\begin{figure*}
\includegraphics[width=\textwidth]{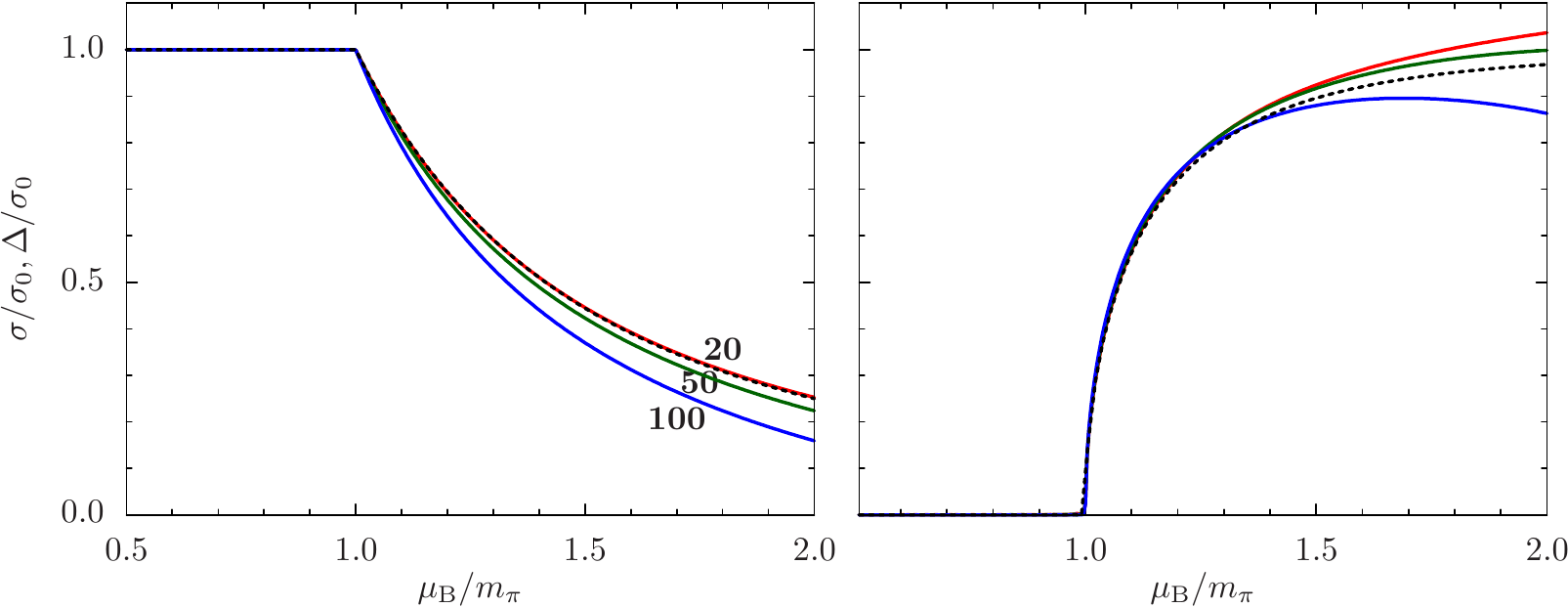}
\caption{Dependence of the rescaled chiral and diquark condensate on $x=\muB/m_\pi$ for several different values of the current quark mass $m_0$. Here $\sigma_0$ is the chiral condensate in the vacuum for given $m_0$ and the other parameters as in Eq.~\eqref{parameters}. The values of $m_0$ in $\text{MeV}$ are indicated in bold (the same color coding is used in both panels). The dashed line shows the prediction of $\chi$PT.}
\label{fig:m0dependence}
\end{figure*}

The first lattice simulations of dense 2cQCD with Wilson quarks were performed with a fixed external source $j$ for the diquark operator~\cite{Hands:2006ve,*Hands:2010gd}; only recently has the extrapolation to vanishing source been studied. However, these attempts were based on fitting three data points by a simple analytical ansatz for the $j$-dependence. Ref.~\cite{Cotter:2012mb} resorted to a linear extrapolation, but as pointed out in Ref.~\cite{Boz:2013rca}, none of the three ans\"atze used therein (linear, power-law, and power-law with an offset) led to satisfactory results.

In principle, adding a diquark source to the NJL model is straightforward and is analogous to introducing the quark mass $m_0$ as a source for the $\bar\psi\psi$ operator. All we need to do is to make the shift $\Delta\to\Delta+j$ in the fermion part of the thermodynamic potential~\eqref{OmegaT0}. However, since we want to gain \emph{analytic} insight into the scaling of the $\Delta$ condensate in the limit $j\to0$, numerical solution of the NJL model is not satisfactory.  Instead we once again employ $\chi$PT. We have confidence in this since the NJL model and $\chi$PT give numerically very similar results in a large range of chemical potentials~\cite{Andersen:2010vu}.

We showed below Eq.~\eqref{LagrangianCHPT} how the diquark source enters the $\chi$PT Lagrangian. Following Ref.~\cite{Kogut:2000ek}, we relate the diquark source to the quark mass by introducing a new angle $\phi$, and express the chemical potential in terms of a dimensionless parameter $x$, via
\begin{equation}
j=m_0\tan\phi\equiv m_0\tilde j,\qquad x\equiv\frac\muB{m_\pi}.
\end{equation}
The static part of the Lagrangian~\eqref{LCHPTstat} with the added diquark source term can then be rewritten as a dimensionless potential,
\begin{equation}
\begin{split}
V(\theta)&\equiv-\frac{\mathscr L_{\chi\text{PT}}^\text{stat}}{4f_\pi^2m_\pi^2}=-\frac12x^2\sin^2\theta-\cos\theta-\tilde j\sin\theta\\
&=-\frac12x^2\sin^2\theta-\frac{\cos(\theta-\phi)}{\cos\phi}.
\end{split}
\end{equation}
Finally introducing the shorthand notation $\delta\equiv\sin\theta$ for the normalized diquark condensate, the stationarity condition $\delta V(\theta)/\delta\theta=0$ takes the simple form,
\begin{equation}
\tilde j=\frac\delta{\sqrt{1-\delta^2}}-x^2\delta.
\label{CHPTgapeq}
\end{equation}
Note that for $\tilde j=0$, we recover the nontrivial solution for the chiral condensate~\eqref{vacuumangle}, now expressed as $\cos\theta=\sqrt{1-\delta^2}=1/x^2$. It is easy to see that for any $\tilde j>0$, Eq.~\eqref{CHPTgapeq} admits a unique positive solution $\delta(\tilde j)$.

\begin{figure*}
\includegraphics[width=\textwidth]{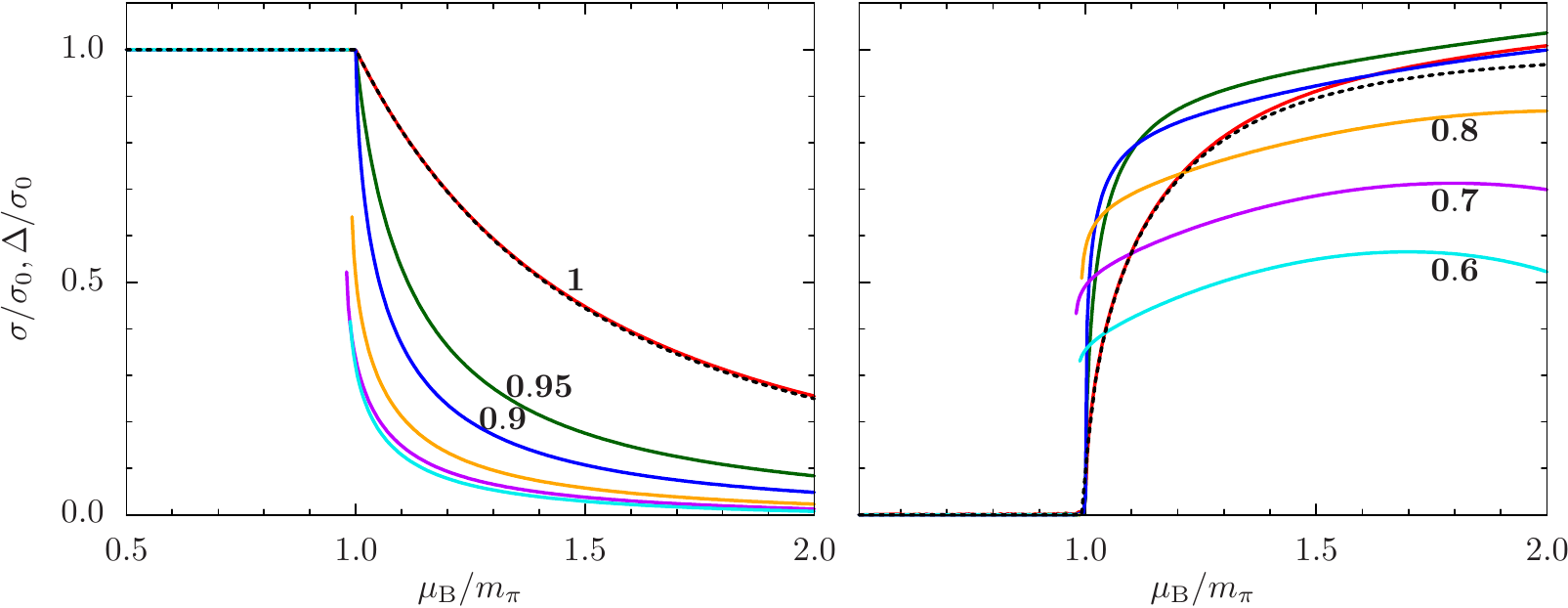}
\caption{Dependence of the rescaled chiral (left panel) and diquark (right panel) condensate on $x=\muB/m_\pi$ for several values of $\lambda$ (shown in bold). Here $\sigma_0\approx300\text{ MeV}$ is the chiral condensate in the vacuum for our parameter set~\eqref{parameters}. The dashed line indicates the prediction of $\chi$PT.}
\label{fig:lambdadependence}
\end{figure*}

The presence of the condensate in the ground state is reflected by a specific asymptotic scaling of this solution in the limit $\tilde j\to0$. The asymptotic expansion of $\delta(\tilde j)$ can be found by an iterative solution of Eq.~\eqref{CHPTgapeq}, leading to
\begin{align}
\notag
\delta(\tilde j)={}&\frac{\tilde j}{1-x^2}-\frac{\tilde j^3}{2(1-x^2)^4}+\mathcal O(\tilde j^5),\qquad(x<1),\\
\notag
\delta(\tilde j)={}&(2\tilde j)^{1/3}-\frac{\tilde j}2+\frac{(2\tilde j)^{5/3}}{24}+\mathcal O(\tilde j^3),\qquad(x=1),\\
\label{asymptoticexp}
\delta(\tilde j)={}&\sqrt{1-\frac1{x^4}}+\frac{\tilde  j}{x^2(x^4-1)}\\
\notag
&-\frac{3x^2\tilde j^2}{2(x^4-1)^{5/2}}+\mathcal O(\tilde j^3),\qquad(x>1).
\end{align}
The $x=1$ part is most easily obtained by writing Eq.~\eqref{CHPTgapeq} as $\tilde j=4\tan^3\frac\theta2/(1-\tan^4\frac\theta2)$, first solving for $\theta(\tilde j)$, and then converting this into a series for $\delta(\tilde j)$. In contrast, the solutions in the regions $x<1$ and $x>1$ are simple Taylor expansions. Moreover, for $x<1$ the expansion only contains odd powers of $\tilde j$, reflecting the unbroken discrete symmetry of Eq.~\eqref{CHPTgapeq}, under which both $\tilde j$ and $\delta$ change sign. We stress that in \emph{both} regions $x<1$ and $x>1$, the condensate is a linear function of the source in the limit $\tilde j\to0$, but the series convergence becomes slower and slower as the phase transition is approached. A rough upper bound on the range of values of $\tilde j$ in which a linear extrapolation makes sense, can be obtained by comparing the first two $\tilde j$-dependent terms of the expansion, giving $\tilde j\lesssim|x^2-1|^{3/2}$ for $x$ close to one in both regions. The convergence of the expansion can also be judged from the exact numerical solution of Eq.~\eqref{CHPTgapeq}, shown in figure~\ref{fig:CHPTsource}.


\subsection{Role of the quark mass}
\label{subsec:qmass}

\begin{figure*}
\includegraphics[width=\textwidth]{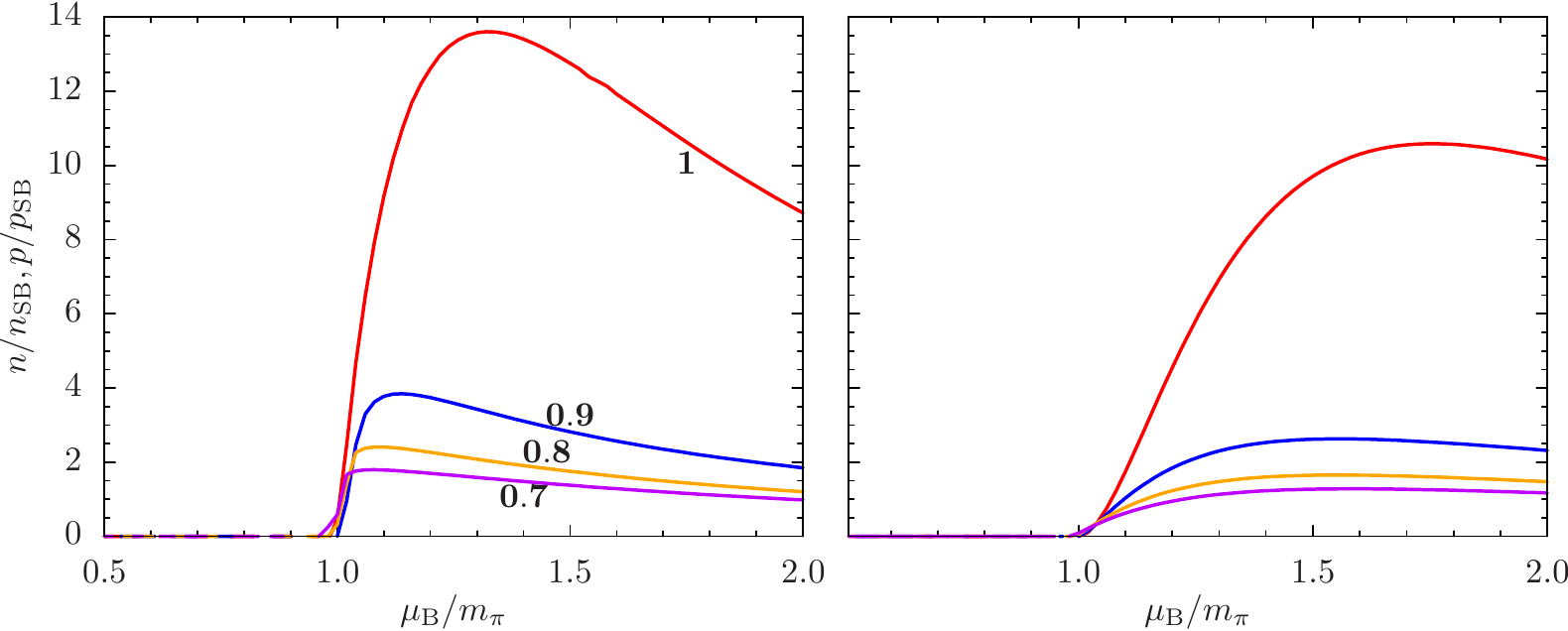}
\caption{Dependence of baryon number density (left panel) and pressure (right panel) on chemical potential for several values of $\lambda$ (shown in bold, the same color coding as in figure~\ref{fig:lambdadependence} is used). Both quantities are normalized to the SB values for a gas of free massless quarks.}
\label{fig:pn}
\end{figure*}

We argued in section~\ref{subsec:puzzle} that reproducing the fast transition to a gas of weakly interacting almost massless quarks at $\muB\geq m_\pi$, as seen on the lattice, requires strong explicit breaking of chiral symmetry. It might be tempting to the think that this effect is due to a large current quark mass. In this section, we therefore set $\lambda=1$ and investigate the dependence on $m_0$ at zero temperature using the mean-field approximation~\eqref{OmegaT0}.

Since the transition to the baryon superfluid phase is expected to occur at $\muB=m_\pi$, it makes sense to trade the chemical potential for the dimensionless parameter $x=\muB/m_\pi$ for the sake of comparison. This is done in figure~\ref{fig:m0dependence}, where the solutions to the gap equations for the chiral and diquark condensates are shown for several values of $m_0$. It is obvious, after proper rescaling, that the prediction of $\chi$PT~\eqref{vacuumangle}, indicated by a dashed line in the figure, works quite well even for unreasonably heavy quarks. (One would probably not expect an approach based on spontaneously broken symmetry to still work even when the mass of the pseudo-Goldstone boson reaches the scale of the ultraviolet cutoff.) It is also clear that just tuning the quark mass not is sufficient for our purposes, even for the largest values of $m_0$: the size of the quasiquark gap $\Delta$ remains largely unaffected~\footnote{The distortion of the curves at high $m_0$ and $\muB$ should be attributed to the three-momentum cutoff.}. Moreover the slight reduction in the chiral condensate is canceled by the increased current quark mass so that the constituent quark mass is not reduced at all.

The results shown in figure~\ref{fig:m0dependence} are also in a good agreement with older lattice simulations using \emph{staggered} quarks~\cite{Hands:2000ei,Hands:2001ee}. There, the prediction~\eqref{vacuumangle} of $\chi$PT was verified numerically for current quark masses varying by an order of magnitude. A tiny reduction of the chiral condensate as compared to Eq.~\eqref{vacuumangle}, observed therein, could --- exactly as in our case --- be ascribed to the relatively large current quark mass. Since the staggered implementation of lattice quarks preserves chiral symmetry, we conclude that heavy quarks alone are not sufficient to explain the thermodynamic behavior in the baryon superfluid phase observed in Ref.~\cite{Cotter:2012mb}. An additional source of chiral symmetry breaking is needed. On the lattice, this is provided by the Wilson term in the action. In our NJL model, the chiral twist serves this purpose, as we will now demonstrate.


\subsection{Role of the chiral twist}
\label{subsec:lambda}

\begin{figure*}
\includegraphics[width=\textwidth]{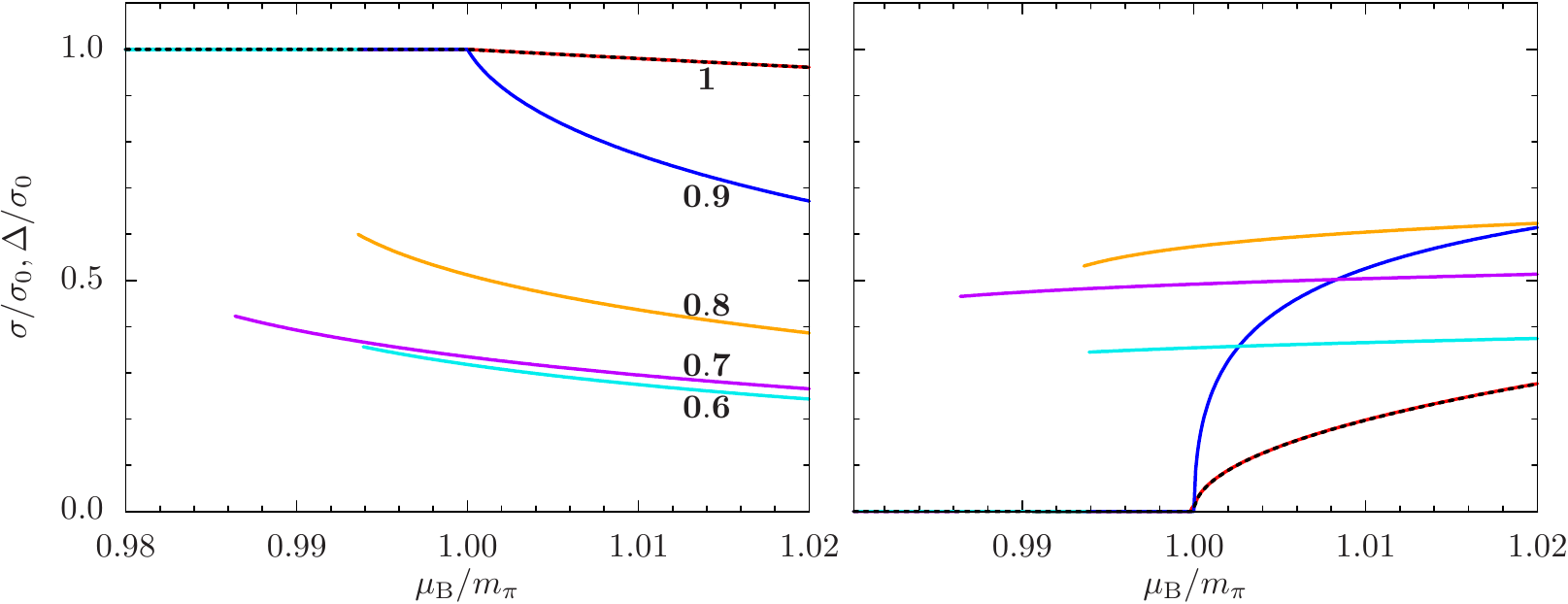}
\caption{Detail of figure~\ref{fig:lambdadependence} close to the phase transition. The same color coding is used.}
\label{fig:lambdadetail}
\end{figure*}

\begin{figure*}
\includegraphics[width=\textwidth]{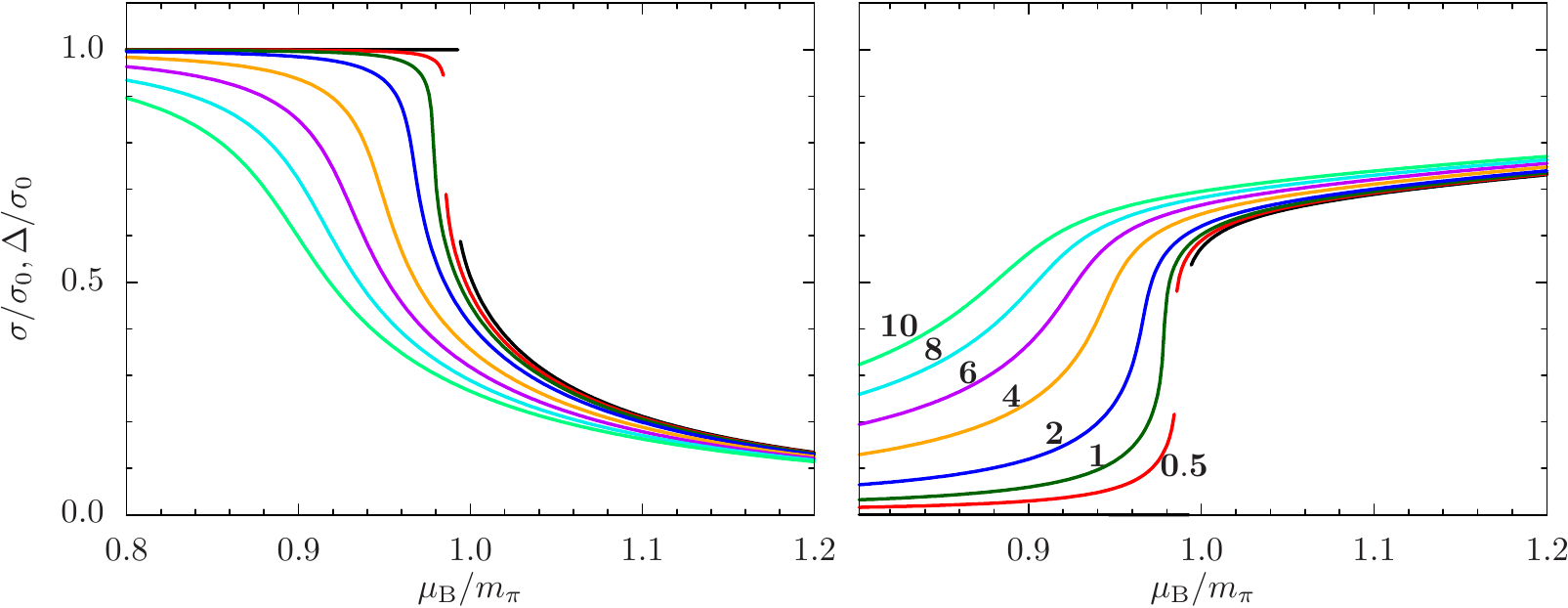}
\caption{Dependence of the rescaled chiral and diquark condensate on the diquark source $j$ for $\lambda=0.8$. The values of $j$ in $\text{MeV}$ are indicated in bold.}
\label{fig:jdependence}
\end{figure*}

In this section, we fix $m_0$ to its value given in Eq.~\eqref{parameters}, and instead vary the chiral twist. Figure~\ref{fig:lambdadependence} shows the dependence of the chiral and diquark condensates at zero temperature on $\lambda$. Since the pion mass is very sensitive to $\lambda$, we again plot these quantities against the ratio $x=\muB/m_\pi$. The numerical results fully confirm our expectations outlined in section~\ref{subsec:modelsetup}, namely that the diquark condensate (at fixed $x$) becomes rather strongly suppressed as $\lambda$ decreases. At the same time, the chiral condensate is strongly suppressed as well. We note that this is in contrast to $\chi$PT where the sum of the two condensates squared is constant. The suppression of both condensates simultaneously is exactly what we want.

How small should $\lambda$ be to reach a quantitative agreement with the lattice results of Refs.~\cite{Cotter:2012mb,Boz:2013rca}? The suppression of the diquark condensate leads to a suppression of the critical temperature $T_d$. In this paper, we are not going to compute the critical temperature, but we can make at least an estimate~\footnote{It would be easy to do in the mean-field approximation, but that would miss an important physical ingredient, namely the order parameter fluctuations. We briefly discuss this point in the conclusions.}. Since in the weak-coupling BCS theory as well as in its extension including the most important corrections due to fluctuations, the critical temperature is proportional to the pairing gap at zero temperature, and since we want $T_d\approx T_c/2$, we need to reduce the diquark condensate roughly by a factor of two. This suggests that $\lambda$ should fall somewhere in the range $[0.6,0.7]$.

Let us next consider the thermodynamic observables baryon number density and pressure. (Since we are at zero temperature, the energy density is linearly dependent on these two via the Gibbs-Duhem relation.) Both quantities are plotted in figure~\ref{fig:pn}. The peak just after the onset of diquark condensation at $\muB=m_\pi$, well formed at $\lambda=1$, is a hallmark of BEC. As $\lambda$ decreases, it is gradually washed away, and again we need $\lambda\in[0.6,0.7]$ to be able to conclude that the BEC-like behavior has given way to BCS-like scaling. One should, however, be aware of the fact that the absolute height of the peak is only indicative here, as it is very sensitive to the value of the pion decay constant and thus to the precise values of the parameters used~\cite{Andersen:2010vu}.

Before closing this section, we would like to point out a subtle detail concerning the dependence of our results on the chiral twist. As may be seen already in figure~\ref{fig:lambdadependence}, the diquark condensation phase transition becomes first order for sufficiently small $\lambda$. In order to highlight this feature, we zoom in on figure~\ref{fig:lambdadependence} in the immediate vicinity of the transition. This is shown in figure~\ref{fig:lambdadetail}. The discontinuity of the condensates becomes rather strong, yet the location of the transition remains very close to $\muB/m_{\pi}=1$. For the values of $\lambda$ considered here, it remains in the range of $\muB/m_\pi\in[0.98,1]$. We have confirmed numerically that around the transition point, the thermodynamic potential has two competing local minima, and used this to locate the transition precisely. The first-order transition only appears for sufficiently low values of $\lambda$, the critical value being approximately $\lambda_c\approx0.88$.

While the appearance of a first-order transition was somewhat unexpected to us, it is not in contradiction with symmetry or any other physical principle. The change of the order of the transition may well be an artifact of our model, and we therefore do not analyze it any further. It is nevertheless interesting to note that introducing the chiral twist provides a mechanism for a direct transition from the vacuum to the BCS regime of weakly coupled quarks without the necessity for an intermediate BEC phase. In any case, we remark that the first-order transition is unlikely to be visible in current lattice simulations, since it appears very close to its expected position, and since even a small external diquark source $j$ is likely to turn it into a crossover. This is clear from figure~\ref{fig:jdependence}, showing details of the transition for $\lambda=0.8$: an external source as small as $1\text{ MeV}$ is sufficient to smooth out the transition into a crossover.


\section{Nonzero temperature: deconfinement}
\label{sec:deconfinement}

At nonzero temperature, the nature of thermal excitations in the system becomes important, and the simple NJL model~\eqref{LagrangianNJL} does not capture the equilibrium thermodynamics of (2c)QCD correctly: at low temperatures, the relevant degrees of freedom of 2cQCD are the pions and the diquarks, whereas the NJL model is based on the quark degrees of freedom. We follow the by now standard procedure and take into account the confining property of strong interactions by coupling the model to the Polyakov loop. We have avoided doing so until now for two reasons: (i) there is considerable freedom in the choice of the gauge sector of the model~\cite{Fukushima:2008wg}, and (ii) the resulting framework is no longer a Lagrangian field theory in the usual sense but merely a statistical model, since there are no dynamical gauge degrees of freedom.

At nonzero temperature, we have to add the effects of thermal quark excitations as well as the gauge sector to Eq.~\eqref{OmegaT0}. The full thermodynamic potential of the PNJL model 2cQCD then becomes~\cite{Brauner:2009gu}
\begin{align}
\label{OmegaPNJL}
&\Omega_\text{PNJL}=\Omega_\text{gauge}(\Phi)+\frac{\sigma^2}{4G}+\frac{|\Delta|^2}{4\lambda G}-2N_c\\
\notag
&\times\sum_\pm\threeint k\Bigl[E^\pm_{\vek k}+T\log\bigl(1+2\Phi e^{-\beta E^\pm_{\vek k}}+e^{-2\beta E^\pm_{\vek k}}\bigr)\Bigr],
\end{align}
where $\Phi$ is the expectation value of the Polyakov loop and $\Omega_\text{gauge}(\Phi)$ is the yet unspecified contribution of the gauge sector.


\subsection{Renormalization of the Polyakov loop}
\label{subsec:PolyakovRenormalization}

Before we proceed with the numerical solution of the model, we have to address a conceptual issue related to the Polyakov loop. On the lattice, the concept of the Polyakov loop is rather subtle as in the naive continuum limit, its expectation value vanishes even in the deconfined phase; the Polyakov loop requires renormalization~\cite{Kaczmarek:2002mc,*Dumitru:2003hp,*Kaczmarek:2005ui}. A simple way to think of it is as follows. The Polyakov loop expectation value can be related to the free energy of a static heavy quark immersed in the colored medium, $F_q$, via $\Phi\sim e^{-\beta F_q}$. An additive renormalization of the free energy then gives a multiplicative renormalization of the Polyakov loop via
\begin{equation}
\Phi_R=e^{-\beta\Delta F_q}\Phi_0,
\end{equation}
where $\Phi_0$ is its ``bare'' value. On the lattice, another interpretation of this formula is available. Using the relation $\beta=N_\tau a_s$, where $N_\tau$ is the number of lattice points in the temporal direction and $a_s$ is the lattice spacing, we have $e^{-\beta\Delta F_q}=(e^{-a_s\Delta F_q})^{N_\tau}\equiv Z_\Phi^{N_\tau}$. The lattice Polyakov loop is a time-ordered product of $N_\tau$ link variables, winding around the temporal direction. Renormalization of each link variable by the constant factor $Z_\Phi$ therefore gives rise to a temperature-dependent renormalization of the Polyakov loop.

In practice, the $Z_\Phi$ factor is found by imposing a certain renormalization condition. Following Refs.~\cite{Cotter:2012mb,Boz:2013rca}, we define this condition by prescribing a value $\bar\Phi_R$ for the renormalized Polyakov loop at a given reference temperature $\bar T$ and $\muB=0$. This leads to the relation
\begin{equation}
\Phi_R(T,\muB)=\Phi_0(T,\muB)\left[\frac{\bar\Phi_R}{\Phi_0(\bar T,0)}\right]^{\bar T/T}.
\label{Polyakovrenorm}
\end{equation}
The $Z_\Phi$ factor is kept constant throughout the computation: it is fixed at $\muB=0$ and the same value is used regardless of the baryon chemical potential.

How should we compare the prediction of our model~\eqref{OmegaPNJL} to the renormalized Polyakov loop measured on the lattice? Note that depending on the reference value $\bar\Phi_R$, the renormalized Polyakov loop can in principle take on any positive value, whereas the quantity $\Phi$ in Eq.~\eqref{OmegaPNJL} is usually assumed (and for some choices of $\Omega_\text{gauge}$ enforced) to lie in the range $[0,1]$. Here we take the point of view that in our model, the finite-valued $\Phi$ is already renormalized, but in some a priori unknown scheme. In order to be able to compare our model to lattice results, we therefore first have to minimize the thermodynamic potential~\eqref{OmegaPNJL} with respect to $\Phi$, and then perform an additional finite renormalization using Eq.~\eqref{Polyakovrenorm} with the same renormalization condition as used on the lattice.


\subsection{Zero chemical potential}
\label{subsec:zeromu}

\begin{figure}
\includegraphics[width=\columnwidth]{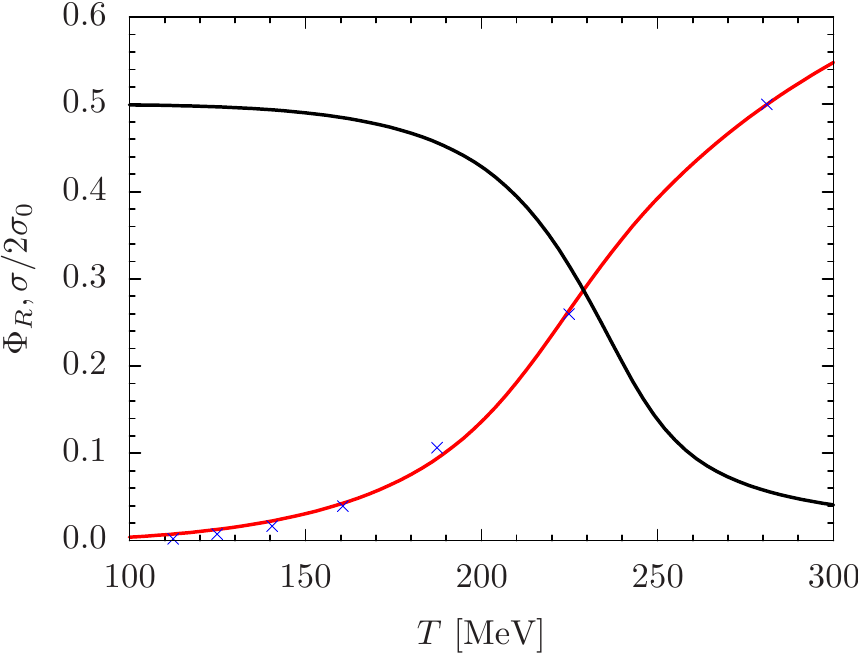}
\caption{Expectation value of the rescaled chiral condensate (black line) and the Polyakov loop (red line) at $\muB=0$ as a function of temperature. The data points are taken from Ref.~\cite{Boz:2013rca}; the highest data point is the reference point which defines the renormalization condition. Note that in order to be able to display both curves on the same scale, the chiral condensate is rescaled to equal $1/2$ in the vacuum.}
\label{fig:zeromu}
\end{figure}

\begin{figure*}
\includegraphics[width=\textwidth]{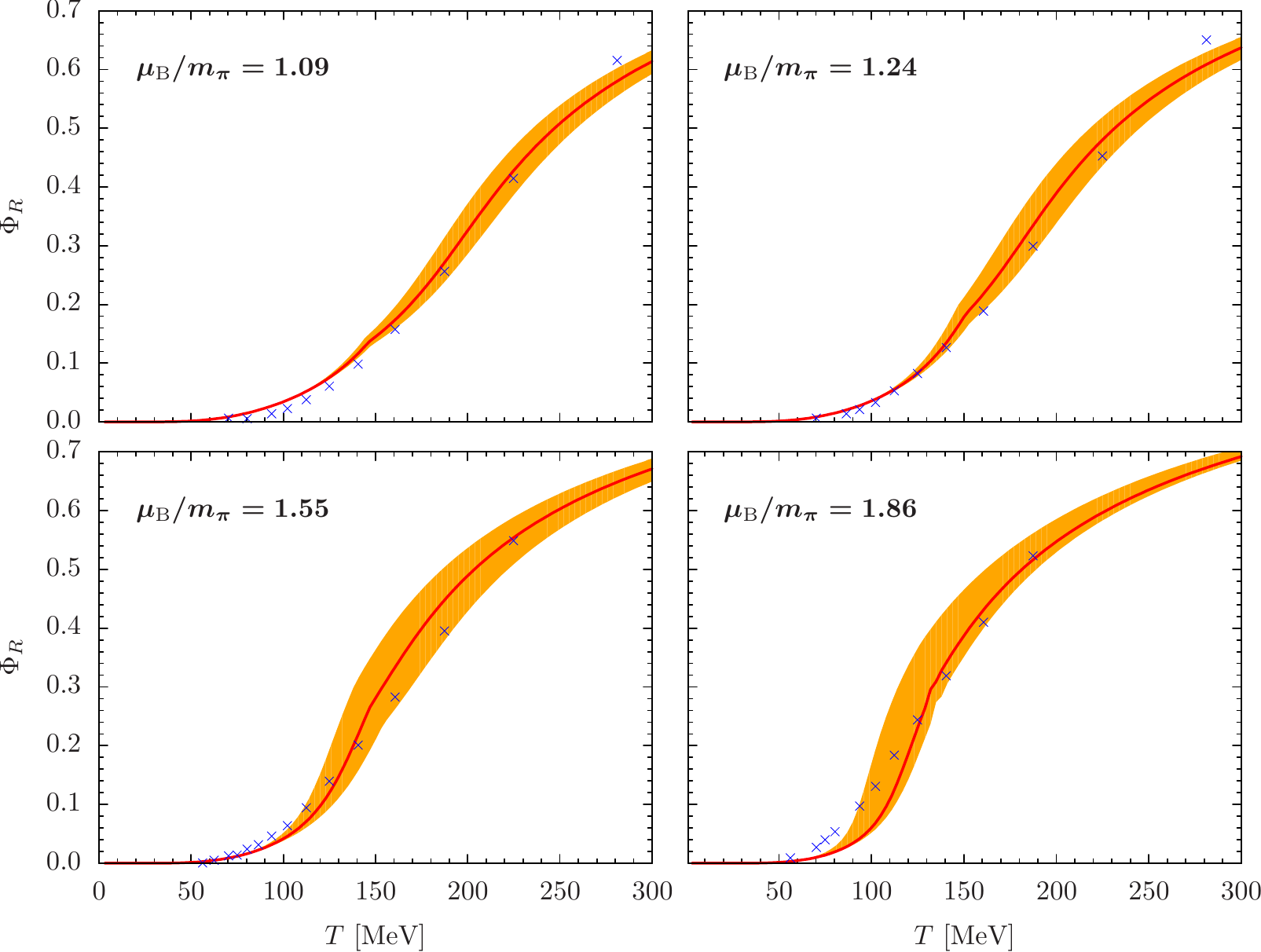}
\caption{Expectation value of the renormalized Polyakov loop for $\lambda=0.7$ and several values of~$\muB$. The red lines represent the best fit of the $T_g$ parameter. The orange bands were obtained by varying $T_g$ in order to estimate the error of the fit. The data points are taken from Ref.~\cite{Boz:2013rca}.}
\label{fig:PolLoopFiniteMu}
\end{figure*}

We next have to make a choice for the Polyakov loop potential $\Omega_\text{gauge}(\Phi)$. Some of the potentials used in the literature include a large number of free parameters, allowing a precise numerical fit to lattice data~\cite{Ratti:2005jh}. However, since we aim at a qualitative understanding rather than numerical fitting, we prefer to have a model with as few free parameters as possible. We therefore employ the potential already used in the context of 2cQCD in Ref.~\cite{Brauner:2009gu}, motivated by the lattice strong-coupling expansion~\cite{Fukushima:2003fw},
\begin{equation}
\Omega_\text{gauge}(\Phi)=-bT\bigl[24\Phi^2e^{-\beta a}+\log(1-\Phi^2)\bigr].
\label{Omegagauge}
\end{equation}
The parameter $a$ is proportional to the deconfinement temperature $T_g$ in the pure-gauge theory via $a=T_g\log24$. The parameter $b$ can be related to the string tension. While the parameters of the quark sector of the model are fixed by Eq.~\eqref{parameters} --- the chiral twist does not affect the mean-field thermodynamic potential at $\muB=0$ --- the parameters $a,b$ will be determined by a fit to lattice data for the expectation value of the Polyakov loop.

To this end, we use the conversion between the lattice and physical units provided by Ref.~\cite{Cotter:2012mb}, defined by the reference temperature of $\bar T=281\text{ MeV}$ for $N_\tau=4$. We then use the renormalization scheme B of Ref.~\cite{Boz:2013rca}, in which $\bar\Phi_R=1/2$. Our best fit to the lattice data is shown in figure~\ref{fig:zeromu} and corresponds to the values
\begin{equation}
b=(278\text{ MeV})^3,\qquad
T_g=247\text{ MeV}.
\label{abcouplings}
\end{equation}
In the same figure, we also display the temperature dependence of the chiral condensate (rescaled for convenience to equal $1/2$ in the vacuum). This demonstrates that with the present values of the gauge sector parameters, differing somewhat from those of Ref.~\cite{Brauner:2009gu}, the chiral and deconfinement crossovers at zero chemical potential appear in the same range of temperatures. Note that the location of the crossover may depend somewhat on the choice of the renormalization condition. We prefer not to give a single value for a pseudocritical temperature due to the ambiguity of this concept.


\subsection{Chemical potential dependence}
\label{subsec:mudependence}

With increasing chemical potential, one generally expects that the deconfinement crossover moves towards lower temperatures due to the back-reaction of the dense medium to the gauge sector. However, it is notoriously difficult to take this back-reaction into account in the PNJL model. A common way around this is to include an explicit $\muB$-dependence in the potential $\Omega_\text{gauge}$, usually using an analytic ansatz motivated by perturbation theory~\cite{Schaefer:2007pw,Herbst:2010rf,*Herbst:2013ail}. One should note that introducing such a chemical potential dependence into $\Omega_\text{gauge}$ leads to an unphysical artifact: it gives an extra contribution to the baryon number density that does not arise from the quark degrees of freedom of the model. However, since $\Omega_\text{gauge}=0$ at $T=0$, this artifact does not violate the ``Silver Blaze'' property stating that at zero temperature, physics is independent of $\muB$ below the onset of diquark BEC~\cite{Cohen:2003kd}.

Here we would like to make the case that the data for the expectation value of the Polyakov loop at nonzero baryon density of Refs.~\cite{Cotter:2012mb,Boz:2013rca}, can be exploited to obtain a direct information about the Polyakov loop potential, without resorting to any ansatz. To make it as simple as possible, we assume that $b$ is constant and fixed by Eq.~\eqref{abcouplings}, while the parameter $a$ contains a $\muB$-dependent temperature scale,
\begin{equation}
a(\muB)=T_g(\muB)\log24.
\end{equation}
This is in accord with the ideas put forward in Ref.~\cite{Schaefer:2007pw}, based on the perturbative running with the physical scale set by the chemical potential. Figure~\ref{fig:PolLoopFiniteMu} shows a comparison of our model predictions with lattice data for different values of $\muB$ taken from Ref.~\cite{Boz:2013rca}. In each plot, we have adjusted the value of $T_g(\muB)$ to represent the lattice data most faithfully; the quality of the fit can be assessed by varying $T_g$, indicated by the bands in figure~\ref{fig:PolLoopFiniteMu}. Along with adjusting $T_g(\muB)$, the $\muB$-independent value $\lambda=0.7$ was chosen to achieve the best overall fit. Note that this value agrees well with the estimate $\lambda\in[0.6,0.7]$ obtained in the previous section, based on different observables for a different thermodynamic regime, namely high density and zero temperature.

\begin{figure}
\includegraphics[width=\columnwidth]{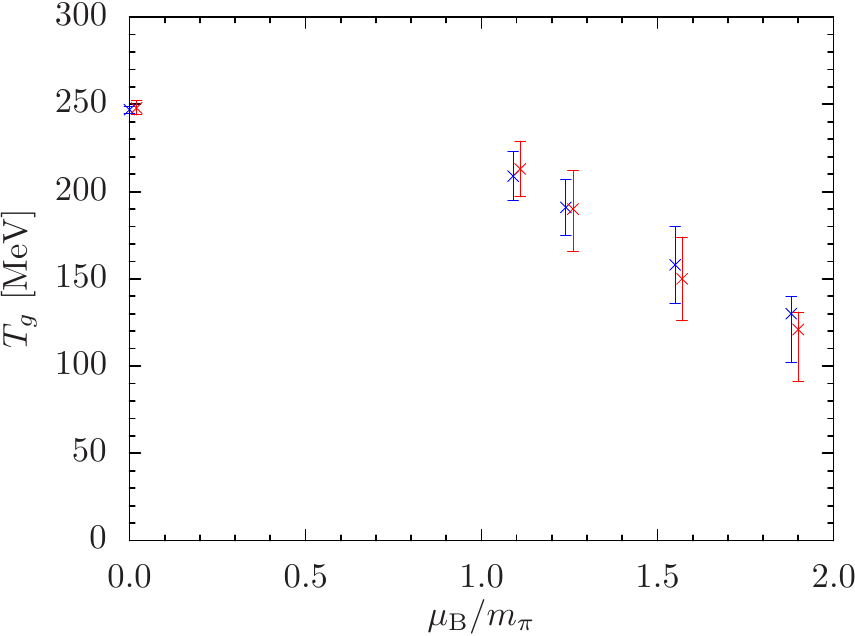}
\caption{Temperature scale $T_g$ of the Polyakov loop potential~\eqref{Omegagauge} as a function of $\muB/m_\pi$. The blue data points are extracted directly from figure~\ref{fig:PolLoopFiniteMu}; the error bars are defined by the bands therein. The red data points correspond to $m_0=48\text{ MeV}$, with the other parameters of the NJL model still given by Eq.~\eqref{parameters}; this reproduces the pion mass used in Ref.~\cite{Boz:2013rca}. Both data sets correspond to the same values of $\muB/m_\pi$, but the red points were slightly displaced for the sake of convenience.}
\label{fig:T0ofM}
\end{figure}

The extracted values of $T_g(\muB)$ are shown as the blue data points in figure~\ref{fig:T0ofM}, with error bars defined by the bands in figure~\ref{fig:PolLoopFiniteMu}. This plot provides a direct information about the back-reaction of the dense medium to the gauge sector, without the need to employ a particular analytical ansatz for the function $T_g(\muB)$. If desired, the data points in figure~\ref{fig:T0ofM} can of course be fitted with a suitably chosen function of $\muB$.

\begin{figure}
\includegraphics[width=\columnwidth]{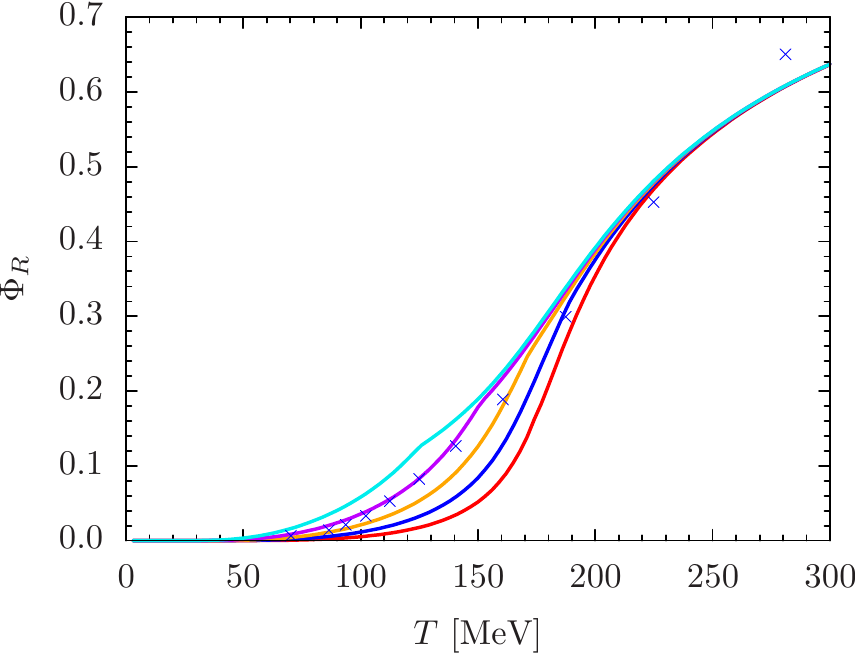}
\caption{Expectation value of the renormalized Polyakov loop for $\muB/m_\pi=1.24$ and several different values of $\lambda$. The same color coding as in figure~\ref{fig:lambdadependence} is used: decreasing $\lambda$ by steps of $0.1$ moves from red ($\lambda=1$) to cyan ($\lambda=0.6$).}
\label{fig:PolLoopLambda}
\end{figure}

The effect of tuning the chiral twist on the Polyakov loop crossover can be appreciated with the help of figure~\ref{fig:PolLoopLambda}. For $\lambda=1$, the crossover tends to be too steep. The reason is that we are in the baryon superfluid phase and the quark gap $\Delta$ is too large to allow thermal excitation of quarks. Hence the behavior of the Polyakov loop to a large extent is the same as in the pure gauge theory. Once $\lambda$ is lowered, the quark gap is reduced and the light quark excitations smear out the crossover. The fact that the curves for different $\lambda$ converge at high temperatures is easy to understand: at these temperatures, the system has already undergone the phase transition to the normal phase (visible in some of the curves as a small cusp) where the thermodynamic potential is independent of $\lambda$ altogether. The residual deviation between different curves is a consequence of rescaling $\muB$ by the $\lambda$-dependent pion mass for the sake of the plot.

Before closing the section, two remarks are in order. Firstly, we have not made an attempt to fit all the parameters of our model precisely to the lattice data. As a consequence, our pion mass (about $550\text{ MeV}$ for $\lambda=0.7$) differs by about 25\% from its lattice value ($720\text{ MeV}$ in physical units). The latter can be reproduced in our model by setting $m_0=48\text{ MeV}$ while keeping the other parameters in Eq.~\eqref{parameters} unchanged. Following the same procedure, that is fitting the parameters $a,b$ at $\muB=0$ to the lattice data and then re-adjusting $T_g$ for each particular value of $\muB$, we arrive at the red data points shown in figure~\ref{fig:T0ofM}. The proximity of the two data sets demonstrates that our results are rather robust and to a large extent independent of the precise model setup. Note that in this latter calculation, including a nonzero diquark source is essential to achieve a good agreement with the lattice data. The effect of the diquark source on the expectation value of the Polyakov loop is much weaker for the light quark parameter set~\eqref{parameters}, and all results shown in figures~\ref{fig:zeromu} and~\ref{fig:PolLoopFiniteMu} are understood at $j=0$.

Secondly, as can be seen in figure~\ref{fig:PolLoopFiniteMu}, our model cannot reproduce very well the Polyakov loop data at both low and high temperatures. Reaching agreement here may require going beyond the simple model of Eq.~\eqref{Omegagauge} and introducing a more phenomenological gauge sector potential with a larger number of free parameters, or using input from other methods~\cite{Haas:2013qwp}. Nevertheless, for lower temperatures \emph{and} lower $\muB$, our simple model can reproduce the lattice data quite well, which makes our conclusions based on symmetry and its explicit breaking rather robust.


\section{Conclusions and outlook}
\label{sec:conclusions}

In the present paper, we have studied in detail the PNJL model with focus on comparing it to lattice data for 2cQCD. We have discussed extensively the effects of strong explicit chiral symmetry breaking due to finite quark masses and the Wilson term in the lattice action. We argued that taking these effects into account requires a generalization of the standard PNJL model. This is accomplished by introducing the chiral twist that incorporates explicit breaking of chiral symmetry in the four-quark interaction term. Based on diverse pieces of evidence, we have argued for its value in the range $\lambda\in[0.6,0.7]$, which can \emph{simultaneously} account for the following nontrivial effects:
\begin{itemize}
\item Rapid crossover from the vacuum to a BCS-like gas of almost massless and gapless quarks at $\muB\gtrsim m_\pi$.
\item Strong suppression of the critical temperature $T_d$ for baryon number breaking as compared to the (pseudo)critical temperature $T_c$ for chiral restoration at $\muB=0$.
\item Strong broadening of the Polyakov loop crossover within the baryon superfluid phase.
\end{itemize}
We argued in section~\ref{subsec:qmass} that these effects cannot be explained by a large current quark mass. However, this statement should be understood \emph{within} the (P)NJL model framework. We cannot discriminate with confidence the effects of explicit chiral symmetry breaking by lattice discretization and by the current quark mass on the level of the 2cQCD Lagrangian. Which of these sources of symmetry breaking is more important can be decided with future lattice simulations by varying the quark mass and lattice spacing.

In addition, we were able to extract the $\muB$-dependent temperature scale $T_g(\muB)$ from the lattice data for the expectation value of the Polyakov loop. This is a model-independent result in the sense that it does not rely on a particular analytical ansatz for the $\muB$-dependence of the Polyakov-loop potential. However, one should not interpret $T_g$ as a deconfinement temperature: while the actual transition from hadronic to quark degrees of freedom is a smooth crossover which does not have a well-defined critical temperature, the quantity $T_g$ is well defined, but depends on the choice of the Polyakov loop potential.

Our simple mean-field model of course has its limitations, most importantly in that it ignores fluctuations of the composite bosonic degrees of freedom.  We therefore have to carefully choose observables which are not sensitive to such fluctuations. (Their effect on the phase diagram of 2cQCD was investigated in Ref.~\cite{Strodthoff:2011tz,*Strodthoff:2013cua}.) Examples of such observables, discussed in this paper, include:
\begin{itemize}
\item[$+$] Thermodynamical observables at low temperature and high density, where the physics is dominated by a Fermi sea of quarks as well as the chiral and diquark condensates. Indeed, it is well-known that the mean-field BCS pairing theory works quantitatively quite well at weak coupling and zero temperature.
\item[$+$] Expectation value of the Polyakov loop from low to high temperatures. The Polyakov loop couples only indirectly to, and therefore is very mildly affected by, the colorless diquark degrees of freedom~\cite{Brauner:2009gu}. The dominant matter contribution to the Polyakov loop thermodynamics comes from the colored quarks, which we do take into account.
\end{itemize}
Observables for which the diquark fluctuations play an essential role, and which we therefore deliberately avoid discussing in this paper, include:
\begin{itemize}
\item[$-$] Critical temperature $T_d$ for baryon superfluidity. This is the main reason why we prefer not to plot the phase diagram of 2cQCD.
\item[$-$] Baryon number density in the confined phase without a diquark condensate (that is, at low $T$ as well as $\muB$).
\end{itemize}

There are several open questions that we would like to understand better and we leave them for future work. Firstly, we have not provided any microscopic derivation of the chiral twist. Both interaction terms in Eq.~\eqref{LagrangianNJL} are expected to arise as we integrate out high-momentum modes in 2cQCD to arrive at a low-energy effective description. In particular, it would be desirable to have an explanation for the fact that $\lambda<1$. We simply assumed this since it leads to the desired phenomenology.

Secondly, we have not touched upon the fact that in the lattice simulations of Ref.~\cite{Cotter:2012mb}, a second onset is observed at high $\muB$, where the thermodynamic quantities rise significantly above their SB limits. At present, we do not have any explanation for this effect.

Finally, we wish to make the case that current and future lattice data for dense two-color QCD should be used to improve our understanding of the real world. To this end, we need a dictionary to translate between models for two-color and three-color QCD. What we have in mind is a framework valid for both two and three colors, augmented by a mapping of the model parameters. For example, one can use the expected scaling based on large-$N_c$ arguments. This was done here and in Ref.~\cite{Brauner:2009gu} to fix the parameters of the NJL part of the model. The gauge sector can be treated in similar manner. Once we have a specific analytical model that reproduces or at least fits the data for $T_g(\muB)$, we can extract an improved $\muB$-dependent Polyakov-loop potential for QCD simply by assuming that its quark-induced part is suppressed by a factor of $1/N_c$ relative to the dominant, gluon contribution. The resulting model will allow one to study, for example, the interplay of chiral symmetry restoration and deconfinement in cold and dense quark matter.


\appendix

\section{Chiral model with a Wilson term}
\label{app:wilson}

In this appendix, we consider the effects of adding a Wilson term to the NJL Lagrangian.


\subsection{Three-momentum cutoff}

Let us revisit the model of Eq.~\eqref{LagrangianNJL}. In the mean-field approximation, the Lagrangian describes a pair of fermion species with squared masses
\begin{equation}
M_\pm^2=\frac1{2\kappa^2}\left(1+2\kappa M\pm\sqrt{1+4\kappa M}\right),
\label{MWilson}
\end{equation}
where $M=m_0+\sigma$. In the limit $\kappa\to0$, one of the masses approaches $M$, as expected. The other, however, diverges as $1/\kappa$. In lattice field theory, this mode is interpreted as the Wilson doubler which decouples in the continuum limit.

We shall now argue that a naive three-momentum cutoff $\Lambda$ is not a suitable regulator in a theory with a heavy doubler. To see this, consider the thermodynamic potential in the vacuum. The thermodynamic potential is given by
\begin{equation}
\Omega^\text{vac}_\text{NJL}=\frac{\sigma^2}{4G}-4N_c\sum_{e=\pm}
\threeint k\,\epsilon_{e\vek k},
\end{equation}
where $\epsilon_{e\vek k}\equiv\sqrt{\vek k^2+M_e^2}$. The gap equation for the chiral condensate is obtained by taking the derivative with respect to $\sigma$. This yields
\begin{equation}
\sigma=\frac{4GN_c}\kappa\sum_{e=\pm}\threeint k\,\frac1{\epsilon_{e\vek k}}\left(1+\frac e{\sqrt{1+4\kappa M}}\right),
\end{equation}
generalizing Eq.~\eqref{condsigma}. The core of the problem is visible now. In the limit $\kappa\to0$, the integrand involving the heavy doubler approaches a nonzero constant. Hence the heavy fermion gives a nonzero contribution to the gap equation even in the limit when its mass goes to infinity. The heavy doubler does not decouple. This can be traced to the fact that the Wilson term modifies the spectrum of the theory, yet the momentum cutoff is sensitive to momenta only, not to masses. We need to use some regularization scheme which affects masses directly. 


\subsection{Pauli-Villars regularization}
\label{subapp:PVvacuum}

Implementing the Pauli-Villars (PV) scheme may be subtle since covariant regulators in general tend to introduce unphysical artifacts at finite temperature and chemical potential. We therefore first derive the naive expression for the thermodynamic potential with the Wilson term, and subsequently introduce the regulator. This 
will ensure thermodynamic consistency.

Returning to Eq.~\eqref{master_action} and switching to Euclidean space, the mean-field thermodynamic potential reads
\begin{widetext}
\begin{equation}
\Omega_\text{PNJL}=\Omega_\text{gauge}(\Phi)+\frac{\sigma^2}{4G}+\frac{|\Delta|^2}{4\lambda G}-2T\sum_n\threeint k\log\det\mathcal G^{-1}(\omega_n,\vek k),
\label{WilsonTDpot}
\end{equation}
where the determinant of the inverse propagator takes the form
\begin{equation}
\det\mathcal G^{-1}=\Bigl\{\bigl[(M-\kappa K_\uparrow^2)^2-K_\uparrow^2\bigr]\bigl[(M-\kappa K_\downarrow^2)^2-K_\downarrow^2\bigr]
+2|\Delta|^2\bigl[(M-\kappa K_\uparrow^2)(M-\kappa K_\downarrow^2)-K_\uparrow\cdot K_\downarrow\bigr]+|\Delta|^4\Bigr\}^2.
\label{determinant}
\end{equation}
\end{widetext}
The arrows indicate momenta of red quarks and green antiquarks, $K^\mu_{\uparrow,\downarrow}\equiv(K^0_{\uparrow,\downarrow},\vek k)$, with
\begin{equation}
K^0_\uparrow\equiv\imag\omega_n+\mu-\imag\alpha,\qquad
K^0_\downarrow\equiv\imag\omega_n-\mu-\imag\alpha.
\end{equation}
Furthermore, $\omega_n\equiv(2n+1)\pi T$ are the fermionic Matsubara frequencies and $\alpha$ is the constant background color gauge field, related to the Polyakov loop variable via $\Phi=\cos(\beta\alpha)$. It is not easy to factorize the determinant~\eqref{determinant} into simple factors corresponding to free fermionic quasiparticles for general $\Delta$. In the following, we therefore set $\Delta=0$ and make sure that $\muB$ is low enough so that we do not enter the baryon superfluid phase. Note that as a consequence $\lambda$ becomes irrelevant for the thermodynamics.

PV regularization is introduced by making the replacement in Eq.~\eqref{WilsonTDpot} 
\begin{equation}
\log\det\mathcal G^{-1}\to\sum_jc_j\log\det\mathcal G_j^{-1},
\end{equation}
where the inverse propagator $\mathcal G_j^{-1}$ is defined analogously to Eq.~\eqref{determinant} by

\begin{equation}
\begin{split}
\det\mathcal G_j^{-1}={}&\Bigl\{\bigl[(M-\kappa K_\uparrow^2)^2+j\Lambda^2-K_\uparrow^2\bigr]\\
&\times\bigl[(M-\kappa K_\downarrow^2)^2+j\Lambda^2-K_\downarrow^2\bigr]\Bigr\}^2.
\end{split}
\label{GPV}
\end{equation}

The coefficients $c_j$ are chosen in order to cancel ultraviolet divergences. Two simple choices, referred to as the PV2 and PV3 schemes, correspond to $j=0,1,2$ with $c_j=\{1,-2,1\}$ and to $j=0,1,2,3$ with $c_j=\{1,-3,3,-1\}$~\cite{Klevansky:1992qe}. The PV2 scheme removes all the divergences, except a residual logarithmic divergence in the thermodynamic potential. This divergence can be eliminated by subtracting the potential at a conveniently chosen reference point. All derivatives of the thermodynamic potential such as the gap equation are rendered finite. The PV3 scheme, on the other hand, renders the thermodynamic potential finite without further subtractions. We choose the PV3 scheme for convenience since the PV2 scheme turns out to be continuous but non-analytic in the limit $\kappa\to0$.

Since at $\Delta=0$ the Dirac determinant~\eqref{determinant} trivially factorizes, it is easy to carry out the Matsubara sum, leading to a generalization of Eq.~\eqref{OmegaPNJL}, valid in the normal phase,
\begin{widetext}
\begin{equation}
\Omega^{\Delta=0}_\text{PNJL}=\Omega_\text{gauge}(\Phi)+\frac{\sigma^2}{4G}-2N_c\sum_{e=\pm}\sum_\pm\threeint k\sum_{j=0}^3c_j\Bigl[E^{j\pm}_{e\vek k}+T\log\bigl(1+2\Phi e^{-\beta E^{j\pm}_{e\vek k}}+e^{-2\beta E^{j\pm}_{e\vek k}}\bigr)\Bigr],
\label{OmegaPNJLkappa}
\end{equation}
where
\begin{equation}
E^{j\pm}_{e\vek k}\equiv\left|\sqrt{\vek k^2+M_{je}^2}\pm\mu\right|,\qquad M^2_{je}\equiv\frac1{2\kappa^2}\left(1+2\kappa M+e\sqrt{1+4\kappa M-4j\kappa^2\Lambda^2}\right).
\label{PVmasses}
\end{equation}
\end{widetext}
The latter generalizes Eq.~\eqref{MWilson} for the Wilson masses to the PV regulator modes. Before we work out the physical consequences, it is useful to pause and explain why we introduced the PV regularization via Eq.~\eqref{GPV}. Here is our consistency check list:
\begin{itemize}
\item The regularization manifestly cancels all divergences. This is easy to see by expanding the logarithm of the regulated determinant~\eqref{GPV} in powers of $\Lambda$. Every factor of $j\Lambda^2$ is suppressed by $1/K^4$ for $\kappa>0$, or at least $1/K^2$ for $\kappa=0$. Since the thermodynamic potential has a quartic divergence, three subtractions are sufficient to make it finite.
\item For $\kappa=0$, our prescription reduces to the simple replacement $M^2\to M^2+j\Lambda^2$, corresponding to the usual implementation of the PV scheme in Lorentz-invariant theories. However, this naive replacement does not work for nonzero $\kappa$.
\item Our thermodynamic potential represents a gas of quasiparticles with masses $M_{je}$ at finite chemical potential. Since the regularization does not affect the shift of $K^0$ by $\mu$, it automatically has the ``Silver Blaze'' property~\cite{Cohen:2003kd}, namely that the physics is completely independent of $\muB$ at $T=0$ below the onset of diquark condensation. Note that this property is not necessarily preserved is some PV-like regularizations where the whole Gram matrix of the Dirac operator is regulated~\cite{Nickel:2008ng}.
\end{itemize}


\subsection{Vacuum physics}
\label{subapp:WilsonThermo}

\begin{figure}
\includegraphics[width=\columnwidth]{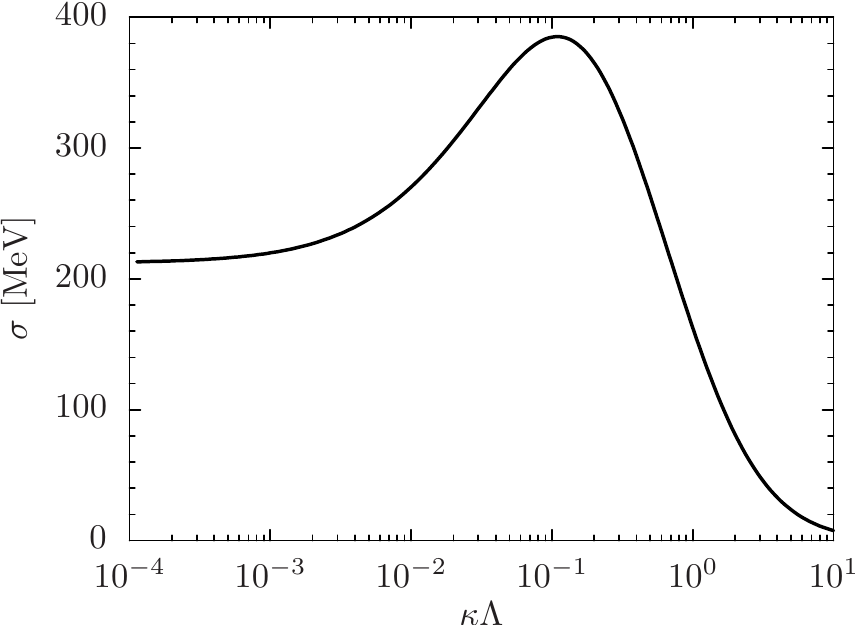}
\caption{Chiral condensate in the vacuum as a function of the dimensionless combination $\kappa\Lambda$; all the other parameters are fixed by Eq.~\eqref{parametersWilson}. Note the logarithmic scale of the horizontal axis.}
\label{fig:wilson}
\end{figure}

As the first step, we have to refit the model parameters owing to the different regularization scheme. The same values for the chiral condensate, pion decay constant, and pion mass given in Eq.~\eqref{observables}, yield the parameter values in the PV3 scheme

\begin{equation}
\begin{split}
G&=5.11\text{ GeV}^{-2},\\
\Lambda&=1129\text{ MeV},\qquad\text{(fitted parameters)}\\
m_0&=5.4\text{ MeV}.
\end{split}
\label{parametersWilson}
\end{equation}

The fact that the same physical observables require a relatively high cutoff and give a comparably small constituent quark mass in covariant regularization schemes is well known~\cite{Klevansky:1992qe}. The Wilson coupling $\kappa$ is treated as a free parameter.

Figure~\ref{fig:wilson} shows the effect of the $\kappa$ coupling on the chiral condensate in the vacuum. As expected, the chiral condensate initially grows with $\kappa$. However, around $\kappa\Lambda\approx10^{-1}$, the growth stops and further increasing $\kappa$ leads to a suppression of the condensate. This is connected to the fact that the PV regulator masses~\eqref{PVmasses} become complex, and we therefore do not attach physical significance to this threshold. Large values of $\kappa$ are not physical anyway since the masses of the two fermion species converge to each other.

In a similar fashion, one can use Eq.~\eqref{OmegaPNJLkappa} to study the consequences of the Wilson term for thermodynamics quantities. However, we do not perform a detailed analysis here, but just remark that it does not affect the qualitative conclusions made in this paper. At a quantitative level, we expect $\kappa$ to play a role whenever chiral symmetry is important. In particular, the presence of an extra heavy fermion species should manifest itself in modified thermodynamics at high temperatures.


\acknowledgments

We are grateful to Jon-Ivar Skullerud for making the lattice data from Ref.~\cite{Boz:2013rca} available to us, and for helpful correspondence. We furthermore appreciate discussions and/or correspondence with Paolo Castorina, Kenji Fukushima and \v Stefan Olejn\'{\i}k. One of us (T.B.) would like to thank the Institute for Theoretical Physics, Goethe University Frankfurt for hospitality during the final stage of this project, and to Michael Buballa, Owe Philipsen, Dirk Rischke, David Scheffler, and Armen Sedrakian for useful comments. The work of T.B.~was supported by the Austrian Science Fund (FWF), grant No.~M 1603-N27.


\bibliography{references}

\end{document}